\documentclass[superscriptaddress, twocolumn, aps, pra, 10pt]{revtex4-2}

\usepackage{graphicx}

\usepackage{amsmath, amssymb, amsfonts}
\usepackage{hyperref}

\usepackage{color}
\begin{document}
     \title{Hybrid Rotational Cavity Optomechanics Using  Atomic Superfluid in a Ring}
    \author{Sanket Das}
        \affiliation{Department of Physics, Indian Institute of Technology Guwahati, Assam 781039, India}
    \author{Pardeep Kumar}
    \email{pardeep.kumar@mpl.mpg.de}
        \affiliation{Max Planck Institute for the Science of Light, Staudtstra\ss{}e 2, 91058 Erlangen, Germany}
    \author{M. Bhattacharya}
        \affiliation{School of Physics and Astronomy, Rochester Institute of Technology, 84 Lomb Memorial Drive, Rochester, New York 14623, USA}
    \author{Tarak N. Dey}
        \affiliation{Department of Physics, Indian Institute of Technology Guwahati, Assam 781039, India}

    \date{\today}

    \begin{abstract}
       We introduce a hybrid optomechanical system containing an annularly trapped Bose-Einstein condensate (BEC) inside an optical cavity driven by Lauguerre-Gaussian (LG) modes. Spiral phase elements serve as the end mirrors of the cavity such that the rear mirror oscillates torsionally about the cavity axis through a clamped support. As described earlier in a related system [\textcolor{blue}{\href{https://doi.org/10.1103/PhysRevLett.127.113601}{P. Kumar et. al., Phys. Rev. Lett. \textbf{127}, 113601 (2021)}}], the condensate atoms interact with the optical cavity modes carrying orbital angular momentum which create two atomic side modes. We observe three peaks in the output noise spectrum corresponding to the atomic side modes and rotating mirror frequencies, respectively. We find that the trapped BEC's rotation reduces quantum fluctuations at the mirror's resonance frequency. We also find that the atomic side modes-cavity coupling and the optorotational coupling can produce bipartite and tripartite entanglements between various constituents of our hybrid system. We reduce the frequency difference between the side modes and the mirror by tuning the drive field's topological charge and the condensate atoms' rotation. When the atomic side modes become degenerate with the mirror, the stationary entanglement between the cavity and the mirror mode diminishes due to the suppression of cooling. Our proposal, which combines atomic superfluid circulation with mechanical rotation, provides a versatile platform for reducing quantum fluctuations and producing macroscopic entanglement with experimentally realizable parameters.
    \end{abstract}

    \maketitle

   \section{\label{sec:level1}Introduction}
    Soon after its experimental realization \cite{doi:10.1126/science.269.5221.198, davis1995bose}, Bose-Einstein condensate (BEC), has emerged as a prominent and controllable system \cite{RevModPhys.73.307} to investigate  and mimic the persistent flow of superfluidity \cite{vilchynskyy2013nature} and superconductivity \cite{doi:10.1126/sciadv.abb9052}. Especially, when confined in multiply-connected geometries (toroidal traps) \cite{RevModPhys.81.647} such systems exhibit persistent currents of superflow \cite{PhysRevA.57.R1505}. These geometries can provide (i) topological protection  to the quantum circulation \cite{Das2012}, (ii) longer dissipationless flow \cite{PhysRevLett.110.025301}, and (iii) `supersonic' rotations \cite{PhysRevLett.124.025301}. Since the first experimental illustration of atomic persistent currents in annularly-trapped BEC \cite{PhysRevLett.99.260401, PhysRevLett.106.130401}, an incredible progress has been made in this configuration to study atomic superflow for the investigation and development of  matter-wave interferometry \cite{PhysRevA.91.013602}, atomtronic circuits \cite{PhysRevLett.111.205301, PhysRevLett.126.170402, Amico2021}, topological excitations \cite{PhysRevLett.100.060401, PhysRevLett.113.135302}, superfluid hydrodynamics \cite{Eckel2014, Wang_2015}, phase slips \cite{PhysRevLett.110.025302, PhysRevA.86.013629}, time crystals \cite{PhysRevLett.123.250402}, gyroscopy \cite{PhysRevA.81.043624} and cosmological studies \cite{PhysRevX.8.021021}.

     Therefore, in a toroidal geometry, it is very important to determine the atomic circulation which involves the phase (winding number) measurement of a rotational state. For detecting the winding number, the current state-of-the-art technologies involve destructive methods \cite{Kumar_2016}, namely, optical absorption imaging of the atoms in the ring which destroys its superfluid character. Furthermore, \textit{in situ} detection in the existing techniques is difficult due to issues related to the optical resolution of the radius of the vortex which demands time-of-flight expansion methods \cite{doi:10.1126/science.1191224}. 
          
     In recent studies, our group proposed a versatile method to detect the magnitude \cite{PhysRevLett.127.113601,Pradhan_PRR2024} as well as direction \cite{Pradhan_PRA2024} of rotation of a bosonic ring condensate with minimal destruction, \textit{in situ} and in times.  Specifically, the method uses the tools of cavity optomechanics \cite{RevModPhys.86.1391}, a unique platform to explore the radiation-pressure interaction of vibrating elements with the photons confined inside an optical resonator. This method not only improves the rotation sensing by three orders of magnitude but also provides a test bed to manipulate the persistent currents by generating the optomechanical entanglement between matter waves. 

    The aforementioned radiation-pressure interaction plays a dual role in cavity optomechanics. On one side, it assists in manipulating the properties of the mechanically pliable objects for applications such as quantum ground state cooling \cite{PhysRevA.77.033804, PhysRevLett.99.093902, PhysRevLett.99.093901} and generation of entanglement between macroscopic objects \cite{PhysRevLett.88.120401, Vitali_2007, PhysRevLett.99.250401, Genes_2008}. On the other end, it can also be used to control the quantum properties of the light. For instance, optomechanical interactions can generate squeezed states of light where the quantum fluctuations in one of the optical quadratures are reduced below the shot-noise level. This comes with increasing fluctuations in other orthogonal quadrature \cite{PhysRevA.49.1337, PhysRevA.49.4055, Qu_2014}. Such engineered squeezed light states play a vital role in enhancing displacement sensitivity in kilometer-sized gravitational wave observatories \citep{Aasi_NatPhoton2013,Ganapathy_PRX2023}, optical communication \cite{Peuntinger_PRL2014}, and metrology \cite{Lee_SciRep2020,Lawrie_ACSPhoton2019}.   

    Interestingly, the rotational analog of cavity optomechanics utilizes radiation torque \cite{Shi_JPB2016} from the angular momentum exchange between the Laguerre-Gaussian (LG) cavity mode and a spiral phase plate as a rotating mirror \cite{Bhattacharya_JOSAB2015}. Such systems have been investigated to cool the rotational degrees of freedom to their quantum ground state \cite {Bhattacharya_PRL2007} and for the realization of opto-rotational entanglement \cite{Bhattacharya_PRA2008,Bhattacharya_PRAII2008}.  
    
   Nowadays, hybrid optomechanical systems \cite{Rogers_QMQM2014,Cernotik_QST2019} pave a versatile pathway in the development of quantum technologies \cite{Barzanjeh2022}. Such systems take into account of a mechanical oscillator interacting with an electromagnetic field \cite{Regal_Nat2008} and an additional physical system \cite{Chauhan_PRA2016} or a degree of freedom \cite{Neukirch_NatPhoton2015}.  In this paper, we present a hybrid setup formed by confining an annularly trapped BEC inside a LG cavity. The spiral phase elements serve as the end mirrors of the cavity such that one mirror rotates about the cavity axis through a clamped support. Specifically, in this hybrid system, we explore (i) ponderomotive squeezing {\it i.e.,} the reduction of quantum fluctuations of the output optical quadratures below the shot noise level at various frequencies, and (ii) the generation of bipartite and tripartite entanglement between the 
    cavity and the matter waves and the macroscopic rotating mirror. The main advantages of our hybrid setup are: (a) From the perspective of toroidal BEC, the atomic rotation in it provides a distinctive tool to correlate the optical amplitude and phase quadratures and provide squeezing of about $87\%$ at a measurement angle of $7^\circ$  around the frequencies of the Bragg-scattered sidemodes. On the other hand, an optimum ponderomotive squeezing of $40$$\%$ below the shot noise level occurs at the angular frequency of the rotating mirror that can further be manipulated by the persistent currents of ring BEC. The same effect also controls the bipartite and tripartite entanglement between the atomic superfluids and the macroscopic object, paving a useful resource for quantum information processing. (b) From the point of view of the LG cavity, it is relatively easy to increase the orbital angular momentum (OAM) of the LG mode. Moreover, in such a cavity setup, the optorotational effects scale with the square of OAM \cite {Bhattacharya_PRL2007} in contrast to the linear scaling of the conventional cavity optomechanics. Using these facts, it is comparatively simple to  manipulate the optical interaction of the ring BEC and the rotating mirror with a common LG mode. This in turn can be harnessed to increase the ponderomotive squeezing and the simultaneous existence of bipartite and tripartite entanglements. Our hybrid system represents a first proposal involving matter wave rotation in hybrid optomechanics and can be exploited for applications in optomechanical sensing, atomtronics and quantum information processing. 
        
    The paper is organized as follows: In Sec. \ref{sec:level2}, we provide details of our hybrid system. In Sec. \ref{sec:level3}, we then provide relevant equations for the quantum dynamics and study the bistability and stability analysis. Sec. \ref{sec:level4} contains a discussion on ponderomotive squeezing while Sec. \ref{sec:entanglement} describes the bipartite and tripartite entanglement in our hybrid setup. Finally, we conclude in Sec. \ref{sec:conc}.   

   \section{\label{sec:level2}Model}
        \label{sec:model}
        The configuration under consideration is a hybrid setup consisting of a ring BEC and an optical cavity. In the following, we provide details of each of these elements. 
 \begin{figure}[h]
 \includegraphics[width=0.45\textwidth]{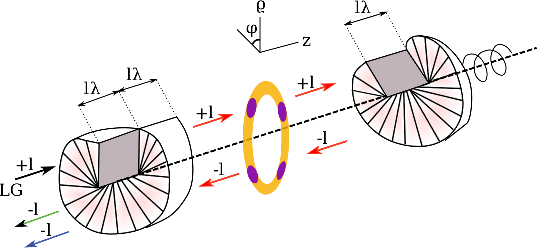}
 \caption{(Color online) Schematic diagram of a hybrid setup consisting of a ring BEC inside a cavity. The cavity comprises two spiral phase elements: one is fixed and partially transmissive, while the other is rotating and highly reflective. The colored arrows show the orbital angular momentum at different positions: The purely extra cavity field, the purely intracavity field, and the intracavity field arising from the intracavity field are green, red, and blue arrows, respectively. The interference between the transmitted and the reflected intracavity fields produces an optical lattice that probes the ring BEC with winding number $L_p$.}
 \label{fig1}
\end{figure}
 \subsection{Ring BEC} 
 The first ingredient of our hybrid setup is an atomic BEC confined in a ring trap and located inside the optical cavity. In the toroidal trap, the harmonic potential experienced by each condensate atom along the radial ($\rho$) and axial ($z$) directions is given by \cite{PhysRevLett.127.113601}
\begin{align}
U(\rho,z)&=\frac{1}{2}m\omega_{\rho}\left(\rho-R\right)^{2}+\frac{1}{2}m\omega_{z}z^{2}\;,\label{eq3}
\end{align} 
where $m$, $\omega_{\rho}$ ($\omega_{z}$) and $R$ represent the mass of the condensate atom, harmonic trapping frequency along the radial (axial) direction and radius of the ring trap, respectively. Due to the above potential, we assume that the atomic evolution along radial and axial directions remains unchanged. However, we consider the dynamical evolution along the azimuthal direction, $\varphi$, as it is not subjected to any potential. This one-dimensional description is possible to be applied within the current state-of-the-art experiments if the number of atoms $N$ of the condensate obeys the following constraint \cite{Morizot_PRA2006}
\begin{align}
N&<\frac{4R}{3a_{\mathrm{Na}}}\sqrt{\frac{\pi\omega_{\rho}}{\omega_{z}}}\;,\label{eq4}
\end{align} 
where $a_{\mathrm{Na}}$ is the $s$-wave scattering length of the sodium atoms in the condensate.
\subsection{Optical Cavity}
 In our setup, an optical cavity is formed with two spiral phase elements with the same handedness. In this arrangement, the input coupler (IC) is a fixed partially transmissive mirror, and the rear mirror (RM) is perfectly reflective. Now, IC is designed to preserve the OAM of the light while transmitting. On the other hand, it removes OAM of $2l\hbar$ per photon in reflection. The perfect reflection from RM removes $2l\hbar$ angular momentum from each photon. In Fig. \ref{fig1}, we provide the mode buildup at the various locations along the optic axis for an input Laguerre-Gaussian beam carrying orbital angular momentum $+l\hbar$. With the above design,  the radiation torque per photon transferred to RM is written as $\xi_{\phi}=2lc\hbar/2L$, where $c$ is the speed of light and $L$ is the cavity length, respectively.
 
\section{\label{sec:level3}Quantum Dynamics}
\subsection{Hamiltonian}
As described above, the modes carrying OAM $\pm l\hbar$ build up inside the cavity, creating an angular lattice about the cavity axis. From such an optical lattice, some of the condensate atoms get coherently Bragg scattered from a macroscopically occupied rotational state with winding number $L_{p}$ to states with $L_{p}\pm 2nl$, with $n=1,2,3,\cdots$. In the following, we consider a dipole potential weaker than the condensate's chemical potential and consider only first-order diffraction, $L_{p}\rightarrow L_{p}\pm 2l$.  

In dimensionless units, the Hamiltonian for our hybrid configuration is expressed as  
\begin{align}
H&=H_{\mathrm{BEC}}+H_{\phi}\;.\label{eq6}
\end{align}   
Here, $H_{\mathrm{BEC}}$ is one-dimensional Hamiltonian for the azimuthal motion of the ring BEC and is governed by  \cite{PhysRevLett.127.113601}
{\color{black}
{
\begin{align}
H_{\mathrm{BEC}}&=\sum_{j = c, d} \Bigg[\frac{\hbar\omega_{j}}{2} \left( X_{j}^{2} + Y_{j}^{2} \right)+\hbar \left(G X_{j}+\frac{U_0 N}{2}\right)a^{\dagger}a\Bigg]\nonumber\\
&+\sum_{j = c, d}2\hbar\tilde{g}N\Big(X_{j}^{2} + Y_{j}^{2}\Big)+2\hbar\tilde{g}N\Big(X_{c}X_{d}-Y_{c}Y_{d}\Big)\;.\label{eq5}
\end{align}
}}
The first term in the square bracket in Eq. (\ref{eq5}) denotes the energies of the Bragg-scattered side modes \cite{PhysRevLett.127.113601} of frequencies $\omega_{c} = \hbar (L_{p} + 2 l)^{2} / (2 I_{a})$ and $\omega_{d} = \hbar (L_{p} - 2 l)^{2} / (2 I_{a})$. The moment of inertia of each atom about the cavity axis is $I_a=m R^2$. {\color{black}{The dimensionless position and momentum quadratures are defined as $X_j=\left(j^\dagger+j\right)/\sqrt{2}$ and $Y_j=\left(j-j^\dagger\right)/i\sqrt{2}$, respectively.}}
 The second term in the square bracket  on the right-hand side of Eq. (\ref{eq5}) governs the effective optomechanical coupling between the side modes and the optical field with the coupling strength  $G = U_{0} \sqrt{N}/{2 \sqrt{2}}$. Here $U_{0}=g_{a}^{2}/\Delta_{a}$ such that $g_{a}$ gives the interaction between single atom and single photon and $\Delta_{a}$ denotes the atomic detuning.  The final two terms arise due to two-body atomic interactions of strength $g=(2\hbar\omega_{\rho} a_{Na})/R$, which can be scaled such that $\tilde{g}= g/4\pi\hbar$. Interestingly,  from the Bogoliubov analysis, the actual side mode frequencies can be written as $\omega_{j}^{\prime} = \sqrt{\omega_{j}^{2} + 4 \omega_{j} \tilde{g} N}$ \cite{PhysRevA.67.013608}. However, for the rest of our analysis, we impose $\omega_{c, d} \gg 4 \tilde{g} N$ and hence use  $\omega_{c, d}^{\prime} \sim \omega_{c, d}$. 

On the other hand, the Hamiltonian for the optical cavity in the rotating frame of the driving laser is governed by
 \begin{align}
 H_{\phi}=-\hbar\Delta_{o}a^{\dagger}a+\frac{\hbar\omega_{\phi}}{2}\left(L_{z}^{2}+\phi^{2}\right)+\hbar g_{\phi}a^{\dagger}a\phi-i\hbar\eta\Big(a-a^{\dagger}\Big)\;,\label{eq1}
\end{align}  
where the first two terms describe the free energy of the detuned cavity mode with $\Delta_0=\omega_l-\omega_0$ and the rotating mirror with resonance frequency $\omega_{\phi}$, respectively. Here $L_z$ and $\phi$ represent the respective dimensionless angular momentum and angular displacement of RM about the cavity axis and these conjugate variables satisfy $[L_z,\phi]=-i$.
 The third term in Eq. (\ref{eq1}) governs the radiation torque on RM, giving rise to optorotational coupling given by
\begin{align}
g_{\phi}&=\frac{cl}{L}\sqrt{\frac{\hbar}{I\omega_{\phi}}}\;.\label{eq2}
\end{align}
The moment of inertia of the rotating mirror about the cavity axis is described as $I=MR_{m}^{2}/2$, where $M$ ($R_{m}$) is the mass (radius) of the RM. 
Finally, the last term on the right-hand side of Eq. (\ref{eq1})represents the cavity drive having amplitude $\eta=\sqrt{P_{in}\gamma_{o}/(\hbar\omega_{o})}$, where $P_{in}$ is the input drive power and $\gamma_{o}$ is the cavity linewidth.

Using Eq. (\ref{eq6}), we derive the Heisenberg equations of motion and include damping and noise appropriately to obtain the following quantum Langevin equations
\begin{align}
\dot{a}-i\left[\tilde{\Delta}-G\left(X_{c}+X_{d}\right)-g_{\phi}\phi\right]a&=-\frac{\gamma_{o}}{2}a+\eta+\sqrt{\gamma_{o}}a_{in}\;,\nonumber\\
\ddot{X}_{c}+\gamma_{m}\dot{X}_{c}+\Omega_{c}^{2}X_{c}&=-\tilde{\omega}_{c}Ga^{\dagger}a-\mathcal{A}X_{d}\nonumber\\
&+\Omega_{c}\epsilon_{c}\;,\nonumber\\
\ddot{X}_{d}+\gamma_{m}\dot{X}_{d}+\Omega_{d}^{2}X_{d}&=-\tilde{\omega}_{d}Ga^{\dagger}a+\mathcal{A}X_{c}\nonumber\\
&+\Omega_{d}\epsilon_{d}\;,\nonumber\\
\ddot{\phi}+\gamma_{\phi}\dot{\phi}+\omega_{\phi}^{2}\phi&=-\omega_{\phi}g_{\phi}a^{\dagger}a+\omega_{\phi}\epsilon_{\phi}\;,\label{eq7}
\end{align}
where $\tilde{\Delta}=\Delta_{o}-\frac{U_{0}N}{2}$ is the effective cavity detuning and $\gamma_{m}$ ($\gamma_{\phi}$) is the damping of each condensate side mode (RM). Further, we have written the quantities $\Omega_{c,d}^{2}=\left(\omega_{c,d}+4\tilde{g}N\right)^{2}-4\tilde{g}^{2}N^{2}$,
$\tilde{\omega}_{c,d}=\omega_{c,d}+2\tilde{g}N$,
$\mathcal{A}=2\tilde{g}N\left(\omega_{c}-\omega_{d}\right)$. In the above Heisenberg-Langevin equation, $a_{in}(t)$ represents the vacuum Gaussian noise of average $\langle a_{in}(t)\rangle=0$, injected into the cavity, and its fluctuations are delta-correlated as
\begin{align}
\label{eq:a_correlation}
\langle \delta a_{in}(t)\delta a_{in}^\dagger(t^{'}) \rangle&=\delta(t-t^{'})\;.
\end{align}
Additionally, $\epsilon_{j}~(j=c,d)$, and $\epsilon_{\phi}$ in the quantum Langevin equation describe the Brownian noise in the condensate side modes and the rotating mirror, respectively. Their average values are zero and fluctuations at respective temperatures $T_{j}$ and $T_{\phi}$ obey following correlations 
\begin{align}
\langle \epsilon_{j}(t)\epsilon_{j}(t{'})\rangle &=\frac{\gamma_{m}}{\Omega_{j}}\int_{-\infty}^{+\infty}\frac{d\omega}{2\pi}e^{-i\omega(t-t{'})}\omega\Big[1+\coth{\frac{\hbar\omega}{2k_{B}T_{j}}}\Big]\;,\label{eq9}\\
\langle \epsilon_{\phi}(t)\epsilon_{\phi}(t{'})\rangle &=\frac{\gamma_{\phi}}{\omega_{\phi}}\int_{-\infty}^{+\infty}\frac{d\omega}{2\pi}e^{-i\omega(t-t{'})}\omega\Big[1+\coth{\frac{\hbar\omega}{2k_{B}T_{\phi}}}\Big]\;,\label{eq10}
\end{align}
where $k_{B}$ is Boltzmann's constant.
\subsection{Steady State}
\begin{figure}[b!]
 \includegraphics[width=0.48\textwidth]{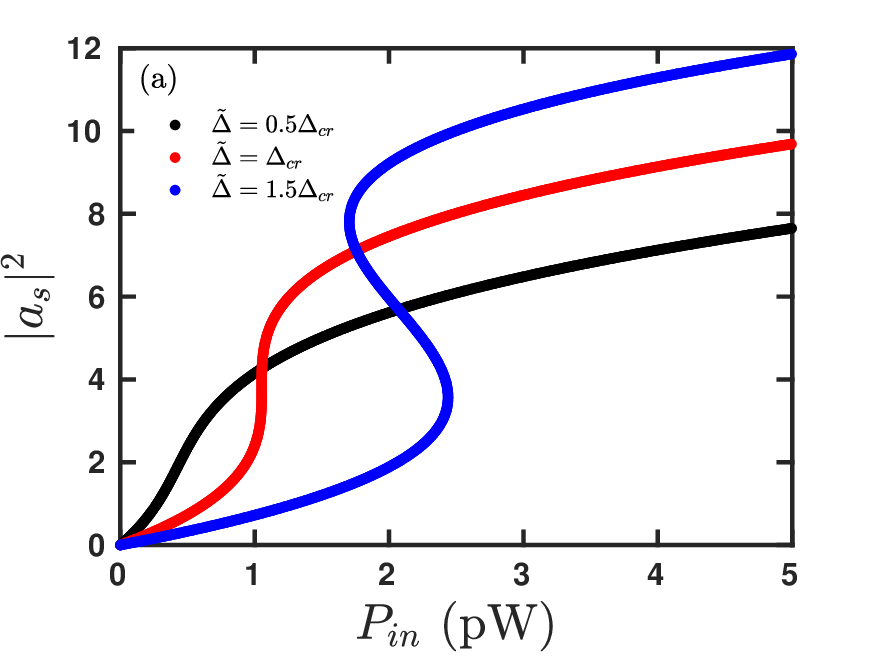}
 \includegraphics[width=0.48\textwidth]{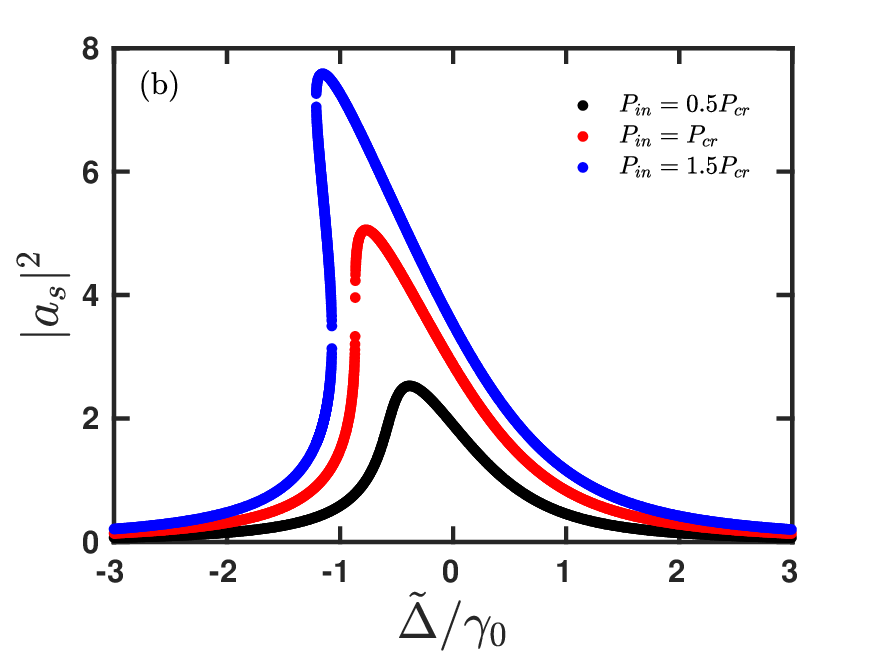}
 \caption{\label{fig:bistability}(Color online) Bistability plot as a function of (a) input power and (b) cavity detuning. Here $\tilde{\Delta}_{\mathrm{cr}}/2\pi=-0.17$ MHz, and $P_{\mathrm{cr}}=1$ pW. The parameters used are $R=12~\mu$m, $N=10^{4}$, $\omega_{\rho}/2\pi=\omega_{z}/2\pi=840$ Hz, $\gamma_{o}/2\pi=0.2$ MHz, $\Delta_{a}/2\pi=5.4$ GHz, $g_{a}/2\pi=0.7$ MHz, $U_{0}/2\pi=90.7$ Hz, $L_{p}=1$, $l=10$, $G/2\pi=3.2$ kHz, $\gamma_{m}/2\pi=0.8$ Hz, $\omega_{o}/2\pi=10^{15}$ Hz, $a_{\mathrm{Na}}=0.1$ nm, $\tilde{g}=14\tilde{g}_{m}$, $\tilde{g}_{m}/2\pi=78.8~\mu$Hz, $L=4$ mm, $M=3.08$ $\mu$g, $\omega_{\phi}/2\pi=653$ Hz, $\gamma_{\phi}=0.1\gamma_{m}$, and $R_{m}=15~\mu$m.}    
 \end{figure}

Following the linearization approach, each operator $\mathcal{O}(t)$ can be
decomposed into its steady-state values $\mathcal{O}_s$ and a small fluctuation $\delta\mathcal{O}(t)$. The steady-state values of each operator are
\begin{subequations}
\begin{align}
a_{s}&=\frac{\eta}{\sqrt{{\Delta^{'}}^{2}+\left(\frac{\gamma_{o}}{2}\right)^{2}}}\;,\\
X_{cs}&=-\tilde{\Omega}_{c}Ga_{s}^{2}\;,\\
X_{ds}&=-\tilde{\Omega}_{d}Ga_{s}^{2}\;,\\
\phi_{s}&=-\frac{g_{\phi}}{\omega_{\phi}}a_{s}^{2}\;,
\end{align}   
\end{subequations}      
{\color{black}where modified cavity detuning is given by $\Delta^{'}=\tilde{\Delta}+\tilde{\Omega}G^{2}a_{s}^{2}+\frac{g_{\phi}^{2}}{\omega_{\phi}}a_{s}^{2}$ and $\tilde{\Omega}=\tilde{\Omega}_{c}+\tilde{\Omega}_{d}$, such that $\tilde{\Omega}_{c,d}=\left({\tilde{\omega}_{c,d}\Omega_{d,c}^{2}\mp\tilde{\omega}_{d,c}\mathcal{A}}\right)/\left({\mathcal{A}^{2}+\Omega_{c}^{2}\Omega_{d}^{2}}\right)$. Here the phase of the cavity drive is chosen such that $a_{s}$ is real. When the effective cavity detuning is larger than the critical value, $\tilde{\Delta}_{cr}=-\sqrt{3}\gamma_0/2$, the above steady-state solution of $|a_s|^2$ manifests a bistable response concerning the input drive field intensity as depicted in Fig. \ref{fig:bistability}(a). Additionally Fig. \ref{fig:bistability}(b) suggests the input drive field intensity ($P_{in}>P_{cr}$) can lead to the bistability response in $|a_s|^2$ where the critical power is $P_{cr}=\hbar\omega_0\gamma_0^2/3\sqrt{3}\left(\tilde\Omega G^2+g_\phi^2/\omega_\phi\right)$. In the remainder of the paper, we choose the parameters to avoid the bistable regime and keep our system monostable. This can also be achieved by using electronic feedback \cite{Schliesser_PRL2006,Arcizet_Nat2006}, which allows us to choose the modified detuning, $\Delta^{'}$, independently of the radiation pressure.}
\subsection{Stability Analysis}
\begin{figure}[h]
\includegraphics[width=0.46\textwidth]{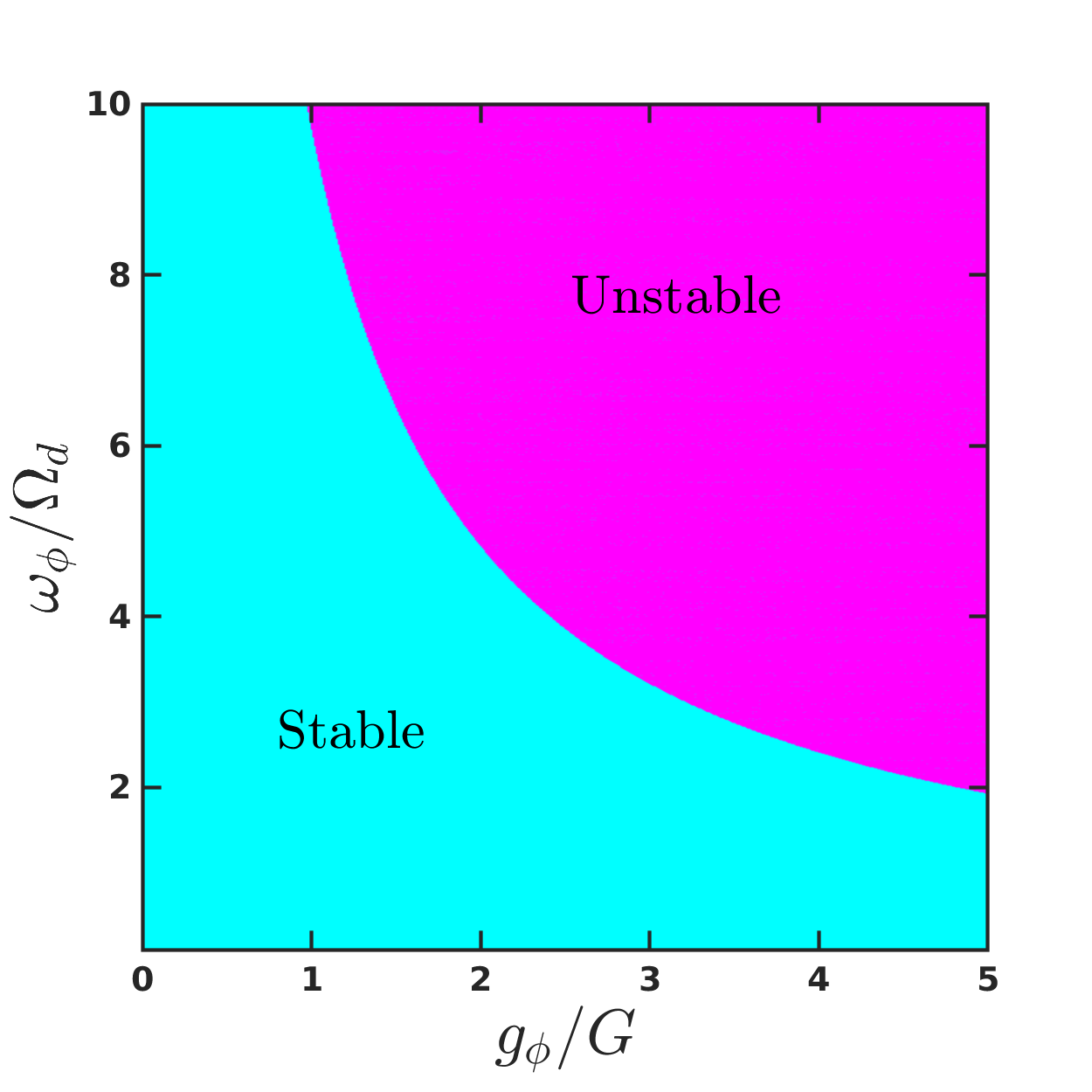}
\caption{(Color online)\label{fig:stability} The stable and unstable regions are determined as functions of the normalized optorotational coupling $g_{\phi}/G$ and the normalized frequency of the rotating mirror $\omega_{\phi}/\omega_d$. We consider $\Delta^{'}=\omega_{\phi}$, $P_{\mathrm{in}}=12.4$ fW. The other parameters are the same as in Fig. \ref{fig:bistability}.}
\end{figure}

The fluctuation part of Eq. (\ref{eq7}) can be expressed as a set of linearized equations
 \begin{align}
 \label{eq:fluctuation_eq}
 \dot{u}(t)&=\tilde{F}u(t)+v(t)\;,
 \end{align}
where the fluctuation vector $u(t)=\left(\delta X_c(t), \delta Y_c(t), \delta X_d(t), \delta Y_d(t), \delta \mathcal{Q}(t), \delta \mathcal{P}(t), \delta \phi(t), \delta L_z(t)\right)^T$ and the input noise vector $v(t)=\left(0,\epsilon_c(t),0, \epsilon_d(t), \sqrt{\gamma_0}\delta \mathcal{Q}_{in}(t), \sqrt{\gamma_0}\delta \mathcal{P}_{in}(t),0, \epsilon_\phi(t)\right)^T$, respectively. We have expressed the cavity field fluctuations in terms of their amplitude and phase quadratures as $\delta \mathcal{Q}=(\delta a+\delta a^{\dagger})/\sqrt{2}$, and $\delta \mathcal{P}=-i\left(\delta a-\delta a^{\dagger}\right)/\sqrt{2}$. The explicit form of the drift matrix $\tilde{F}$ is given by
\begin{align}
\tilde{F}&=\begin{pmatrix}
0 & \Omega_{c} & 0 & 0 & 0 & 0 & 0 & 0\\
-\Omega_{c} & -\gamma_{m} & -\frac{\mathcal{A}}{\Omega_{c}} & 0 & -\frac{\tilde{\omega}_{c}G_{r}}{\Omega_{c}} & 0 & 0 & 0\\
0 & 0 & 0 & \Omega_{d} & 0 & 0 & 0 & 0\\
 \frac{\mathcal{A}}{\Omega_{d}} & 0 &-\Omega_{d} & -\gamma_{m} & -\frac{\tilde{\omega}_{d}G_{r}}{\Omega_{d}} & 0 & 0 & 0\\
  0 & 0 & 0 & 0 & -\frac{\gamma_{o}}{2} & -\Delta^{\prime} & 0 & 0\\
-G_{r} & 0 & -G_{r} & 0 & \Delta^{\prime} &-\frac{\gamma_{o}}{2} & -g_{\phi}^{r} & 0\\
0 & 0 & 0 & 0 & 0 & 0 & 0 & \omega_{\phi}\\
0 & 0 & 0 & 0 &-g_{\phi}^{r} & 0 & -\omega_{\phi} & -\gamma_{\phi}
\end{pmatrix}\;.\label{eq13}
\end{align}
The enhanced side modes-cavity coupling strength and optorotational coupling are described as $G_{r}=\sqrt{2}Ga_{s}$, and $g_{\phi}^{r}=\sqrt{2}g_{\phi}a_{s}$, respectively. We derive the stability condition for the hybrid rotational cavity optomechanical system by invoking the Routh-Hurwitz criterion \cite{Edmund_PRA1987}, which suggests all the eigenvalues of $\tilde{F}$ must have negative real parts. From Fig. \ref{fig:stability}, it is evident that our system lies in the stable region for the weaker single photon optorotational coupling strength with respect to the side mode-cavity coupling strength. Moreover, increasing $g_\phi$ results gradual decrease in the characteristics frequency bound of the RM.
\section{\label{sec:level4}{Ponderomotive Squeezing}} 
In this section, we analyze a hybrid system to manipulate the quantum properties of the output light. Our interest, in particular, is to reduce the quantum fluctuations in the optical quadratures well below the shot noise level and to describe the influence of quantum circulation on the optical squeezing.

\subsection{Quadrature Noise Spectrum}
To describe the ponderomotive squeezing, we invoke the homodyne measurement technique and obtain the quadrature noise spectrum of the output optical field. The homodyne-detected signal can be expressed as \cite{Mancini_PRA1994,Xunnong_PRA2014}
   \begin{align}
   \label{eq:Homodyne}
   \mathcal{Q}^{out}_{\theta}(\omega)&=\mathcal{Q}_{out}(\omega)\cos\theta+\mathcal{P}_{out}(\omega)\sin\theta\;, 
   \end{align}
   where $\theta$ determines the measured field quadrature and is adjusted experimentally.
   The cavity relates the output and input field quadratures as 
    $\mathcal{Q}_{out}(\omega)=\sqrt{\gamma_{o}}\delta \mathcal{Q}(\omega)-\mathcal{Q}_{in}(\omega)$, and  $\mathcal{P}_{out}(\omega)=\sqrt{\gamma_{o}}\delta \mathcal{P}(\omega)-\mathcal{P}_{in}(\omega)$. 
    The output quadrature noise spectrum is then calculated as
    \begin{align}
   \label{eq:psd_definition}
   	 S(\omega)&=|\xi_{1}(\omega)|^{2}+|\xi_{2}(\omega)|^{2}+i\left[\xi^{\ast}_{1}(\omega)\xi_{2}(\omega)-\xi^{\ast}_{2}(\omega)\xi_{1}(\omega)\right]\nonumber\\
   	  &-2\gamma_{m}\omega\Bigg[\frac{|\xi_{3}(\omega)|^{2}}{\Omega_{c}}+\frac{|\xi_{4}(\omega)|^{2}}{\Omega_{d}}\Bigg]\Bigg[1-\coth\Big(\frac{\hbar\omega}{2k_{B}T}\Big)\Bigg]\nonumber\\
   	  &-2\frac{\gamma_{\phi}\omega}{\omega_{\phi}}|\xi_{5}(\omega)|^{2}\Bigg[1-\coth\Big(\frac{\hbar\omega}{2k_{B}T_{\phi}}\Big)\Bigg]\;.
   \end{align}
A full derivation of the above spectrum and detailed expressions of $\xi_{i}$'s are given in Appendix \ref{AppendixA}. Note that the noise spectrum in Eq. (\ref{eq:psd_definition}) is normalized in such a way that $S(\omega)=1$ represents a shot-noise level \cite{Fabre_PRA1994}.
\begin{figure}[ht!]
 \includegraphics[width=0.46\textwidth]{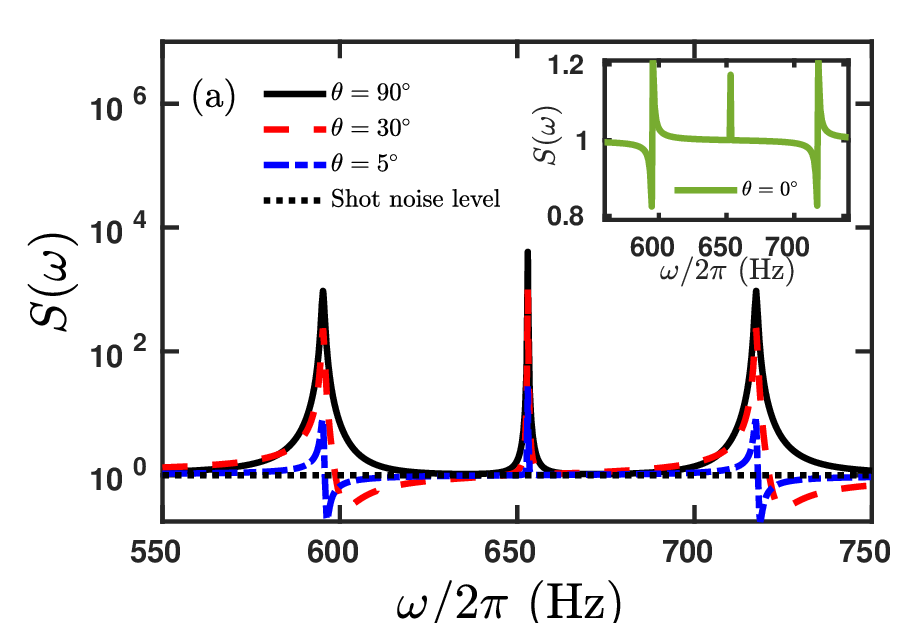}
 \includegraphics[width=0.48\textwidth]{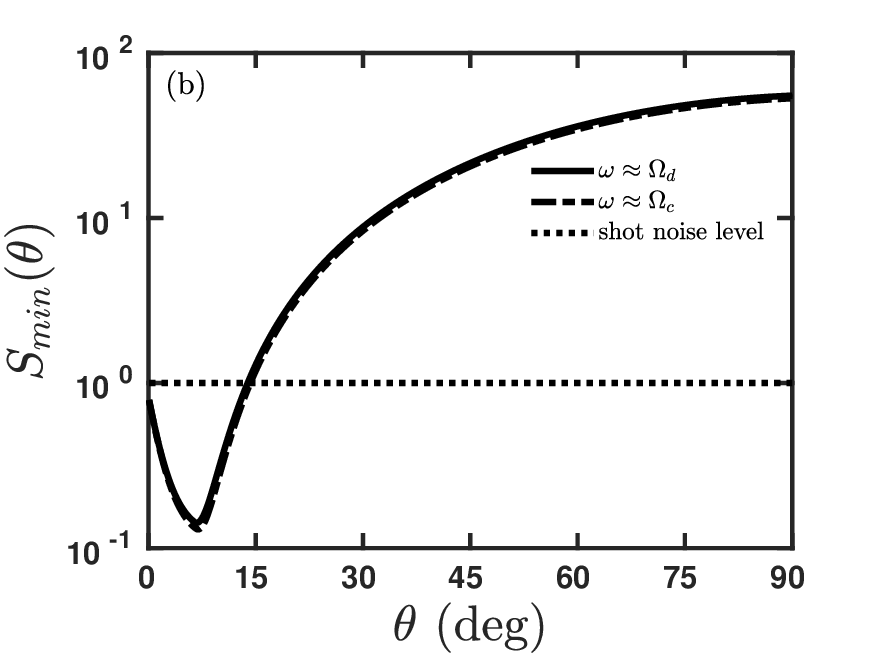}
 \caption{(Color online)\label{fig:psd} (a) Power spectral density (PSD) of the output optical quadrature as a function of the response frequency $\omega/2\pi$,  for the measurement angles $\theta=90^{\circ}$ (black solid line), $\theta=30^{\circ}$ (red dashed line), and $\theta=5^{\circ}$ (blue dot-dashed line). (b) PSDs as a function of the measurement angle $\theta$. Solid red, dashed blue and solid green curves are produced by fixing the response frequency near $\Omega_{c}$, $\Omega_{d}$ and $\omega_{\phi}$, respectively. Also, in these plots the spectrum is normalized to the shot noise (black dotted line). Parameters used are $T=10$ nK, $T_{\phi}=1$ mK and other parameters are same as in Fig. \ref{fig:stability}.}
\end{figure}

In Fig. \ref{fig:psd}, we show the power spectral density (PSD) of the optical quadrature as a function of the response frequency for different homodyne measurement angles. Here the black dotted curve represents the shot noise level. For $\theta=90^{\circ}$, the phase quadrature of the output optical field results in three Lorentzian peaks on top of the shot noise background  at the characteristic frequencies,  $\Omega_{d}/2\pi\sim 595$ Hz, and  $\Omega_{c}/2\pi\sim 717$ Hz corresponding to side modes of the rotating BEC, and the rotating mirror, $\omega_{\phi}/2\pi\sim 653$ Hz, respectively. It is evident that the fluctuations remain above the vacuum noise in this scenario. However, decreasing the measurement angle $\theta$ reduces the value of $S(\omega)$ well below the shot noise level near the atomic side mode frequencies as presented by the blue dot-dashed curve of Fig. \ref{fig:psd}(a). It is a clear signature of the ponderomotive squeezing which occurs
due to the existence of correlations between optical amplitude and phase quadratures.
{\color{black} For instance, at a measurement angle, $\theta=5^\circ$, the output optical noise is reduced by $84\%$ below the shot noise within a bandwidth of $20$ Hz around $\omega=\Omega_c,\Omega_d$ and the amplitude quadratures become asymmetric like the Fano lineshape \cite{Militaru_PRL2022}.} Such a line profile arises due to the interference effects generated by the optical quadratures and the resonant process produced by the atomic density modulation driven by amplitude quadrature \cite{Magrini_PRL2022}. In Fig. \ref{fig:psd}(a), we set the effective cavity detuning equal to $\omega_{\phi}$. As a result, the correlation between amplitude and phase quadratures of the input field produces a small squeezing at $\theta=0^\circ$ \cite{Xunnong_PRA2014}, as shown in the inset of Fig. \ref{fig:psd}(a). 

As discussed above, the measurement angle during the homodyne detection plays a crucial role to reduce the spectral noise below the shot noise. Now, in Fig. \ref{fig:psd}(b), we display the PSD by tuning the measurement angle in the range $[0,180^{\circ}]$ by fixing the response frequencies around the mechanical side modes . For our parameters, the maximum noise reduction near the mechanical side mode frequencies occurs around $\theta=7^{\circ}$ and the maximum of $87\%$ ponderomotive squeezing is obtained.  

\subsection{Optimum Squeezing}
In the preceding section, we have described that in our hybrid model, ponderomotive squeezing can be generated by choosing appropriate measurement angle. This happens only around the frequencies of the mechanical side modes. However, the fluctuations in the quadrature spectrum at the frequency of the rotating mirror still remains well above the background noise floor. {\color{black}In order to reduce the fluctuations,  we optimize the optical quadrature squeezing spectrum $S_{opt}(\omega)$ by choosing $\theta(\omega)$ in such a way that $d S(\omega)/d\theta=0$ for all the frequencies \cite{Mancini_PRA1994}.} This gives
\begin{align}
\label{eq:optimized_theta}
\theta_{opt}(\omega)=\frac{1}{2}\arctan\left[-\frac{B_{2}(\omega)}{B_{1}(\omega)}\right]\;,
\end{align}
where the expressions of $B_{1}(\omega)$ and $B_{2}(\omega)$ are too involved and are shown in the appendix \ref{AppendixA}.
\begin{figure}[h!]
 \includegraphics[width=0.46\textwidth]{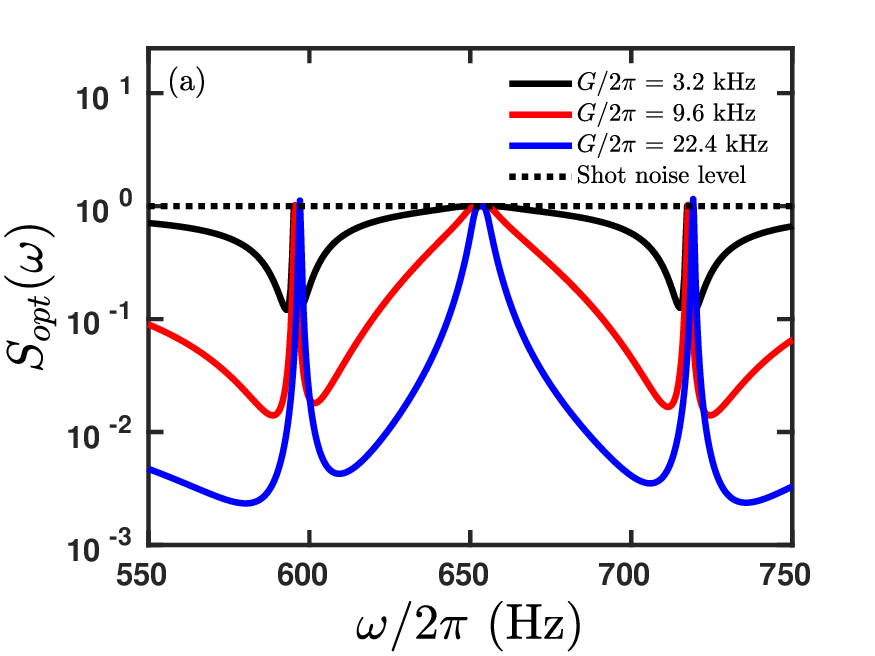}
\includegraphics[width=0.46\textwidth]{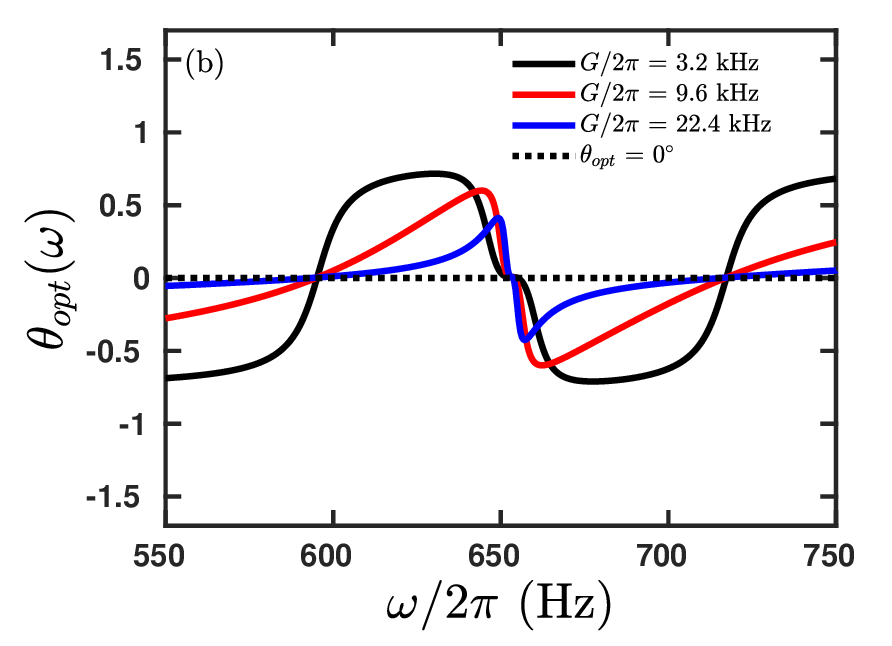}
\caption{\label{fig:optimum_psd_freq}(Color online) Optimized PSD of the optical quadrature as a function of the response frequency for $G/2\pi=3.2$ kHz (black solid line), $G/2\pi=9.6$ kHz (red solid line), and $G/2\pi=22.4$ kHz (blue solid line). Parameters used are same as in Fig. \ref{fig:bistability}.}
\end{figure}
{\color{black}The optimized squeezing spectrum is presented in Fig. \ref{fig:optimum_psd_freq}. The exciting findings of the optimized squeezing spectrum are: (i) there is a significant reduction in the fluctuations below the shot noise level at the vicinity of the side mode frequencies $\Omega_{c,d}$. (ii) Increasing the coupling strength between the atomic side modes with the cavity enhances the ponderomotive squeezing near the atomic side mode frequencies. Further, we observe a significant enhancement in the optical squeezing spectrum at the side mode frequencies $\Omega_{c,d}$. To understand its reason, we plot the optimized homodyne angle as a function of the frequency as depicted in Fig. \ref{fig:optimum_psd_freq}(b). The black-solid, red-solid, and blue-solid curves suggest the optimized homodyne angle becomes zero at the atomic side mode frequencies and at the mirror frequency. At these specific frequencies, the major contribution to the optimized optical squeezing spectrum originates from the first term of Eq. \ref{eq:psd_definition} ($|\xi_1(\omega)|^2$), only. The vanishingly small functional values of $\tilde{F_2}$ (Eq. \ref{eq:defininition_xi}) at $\Omega_c,~\Omega_d,~\textrm{and}~\omega_\phi$ lead the optimal squeezing to the unity.} 

\begin{figure}[h!]
 \includegraphics[width=0.46\textwidth]{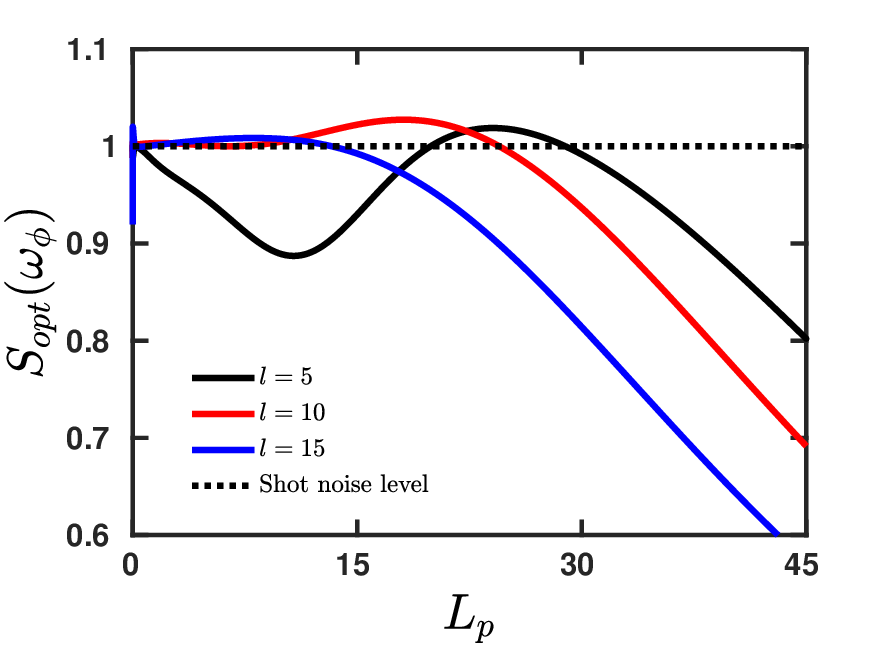}
 \caption{\label{fig:optimum_psd_freq_Lp}(Color online) Optimized PSD of the output optical quadrature plotted at the frequency of the rotating mirror as a function of the winding number $L_p$. Parameters used are same as in Fig. \ref{fig:bistability}.}
 \end{figure}      
So far we have described the ponderomotive squeezing in our hybrid system by tuning various parameters. However,  such squeezing occurs only at the frequencies of the Bragg-scattered mechanical side modes, whereas the optical quadrature fluctuations at the frequency of rotating mirror reduced just to the level of shot noise. In Fig. \ref{fig:optimum_psd_freq_Lp}, we explore the influence of the persistent flow in the ring BEC to manipulate the noise reduction and to achieve ponderomotive squeezing at the resonance frequency of the rotating mirror. The blue solid curve of Fig. \ref{fig:optimum_psd_freq_Lp}, suggests the increase in the winding number of the BEC enhances its interaction with the OAM carrying input field ($l=15$) to obtain $40\%$ of ponderomotive squeezing. A weaker topological charge of the input field requires a relatively higher winding number of the rotating BEC to produce a similar amount of ponderomotive squeezing. Additionally, for lower $L_p$ values, the atomic collisions dominate to give rise to the optical mode squeezing. 

\section{Entanglement} 
\label{sec:entanglement}
{\color{black}
In the preceding section, we have exploited the radiation pressure force to squeeze the quantum fluctuations of the output light field. The radiation pressure also plays a crucial role in cooling down the rotational mirror to its quantum ground state and producing entanglement. In particular, our hybrid system sets a stage where the interactions between the atomic side modes with the optical field and the radiation torque play a pivotal role in obtaining the bipartite entanglements between various degrees of freedom. 
\subsection{Bipartite Entanglement}
To quantify the entanglement between various subsystems, we use the linearized dynamics of Eq.(\ref{eq:fluctuation_eq}) and the Gaussian character of the quantum noise to achieve the stationary Gaussian state, which can be fully characterized by a $8\times 8$ covariance matrix $V$, whose elements are written as $V_{ij}=\frac{1}{2}\langle u_{i}(\infty)u_{j}(\infty)+u_{j}(\infty)u_{i}(\infty)\rangle$. Under the stable condition, the covariance matrix $V$ satisfies the Lyapunov equation \cite{RevModPhys.86.1391}
\begin{align}
\label{eq:lyapunov}
\tilde{F}V+V\tilde{F}^{T}=-D,
\end{align}
where the matrix of the stationary noise correlation function is $D=\textrm{diag}\{0,\gamma_m(2n_c+1),0,\gamma_m(2n_d+1),\gamma_0/2,\gamma_0/2,0,\gamma_\phi(2n_m+1)\}$. The mean thermal excitation for the BEC side modes and the mechanical excitations of the rotating mirror are $n_i = \left(\textrm{exp}\{\hbar\Omega_i/k_B T\}-1\right)^{-1}~(i=c,d)$,  and $n_m = \left(\textrm{exp}\{\hbar\omega_\phi/k_B T_\phi\}-1\right)^{-1}$, respectively. Using the above formalism, we
study the two body entanglement in the hybrid system by evaluating the logarithmic negativity $E$, defined as \cite{Ghobadi_PRA2011}
\begin{align}
\label{eq:log_negativity}
E&=\max[0,-\ln 2\eta^{-}]\;,
\end{align}
 where $\eta^{-}=2^{-1/2}[{\Sigma(V)-[\Sigma(V)^2-4\det (V_{sub})]^{1/2}}]^{1/2}$, with $\Sigma(V)=\det A + \det B - 2\det C$. Here, $V_{sub}$ is a generic $4\times 4$ submatrix 
\begin{align}
V_{sub}&=\begin{pmatrix}
A & C\\
C^T & B\\
\end{pmatrix}\;,
\end{align}
representing a particular bipartite system under consideration. $A,B$ and $C$ are $2\times 2$ blocks of the covariance matrix. The bipartite entanglement exists if $E>0$ {\it i.e.,} when $\eta^{-}<1/2$. 
\begin{figure}[h!]
	\includegraphics[width=0.42\textwidth]{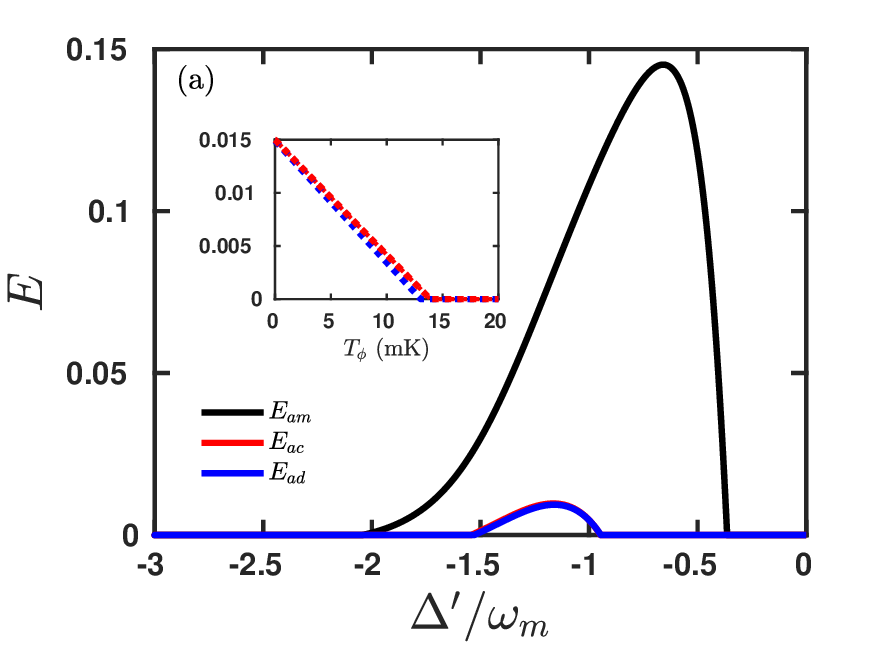}
	\includegraphics[width=0.42\textwidth]{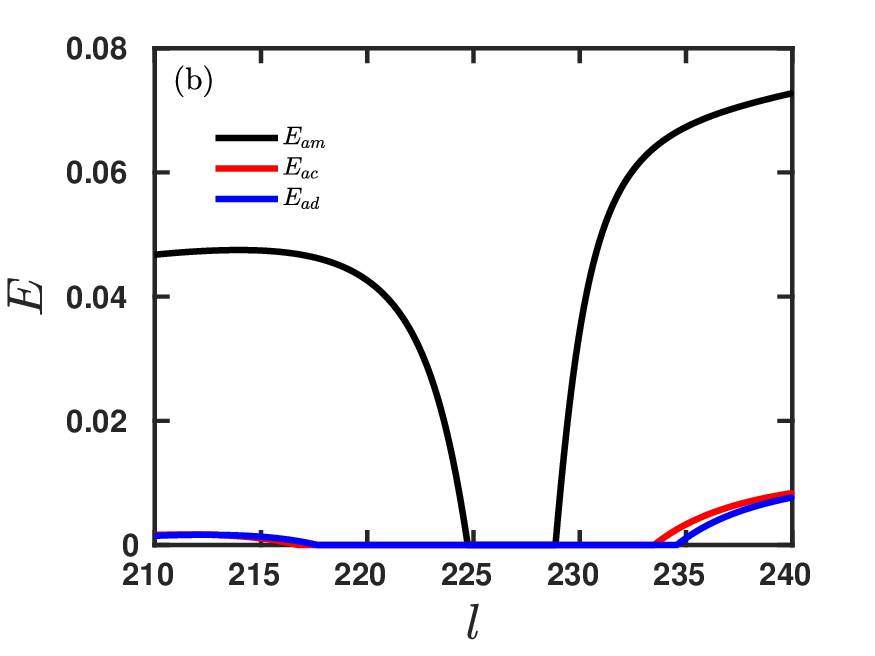}
    \includegraphics[width=0.42\textwidth]{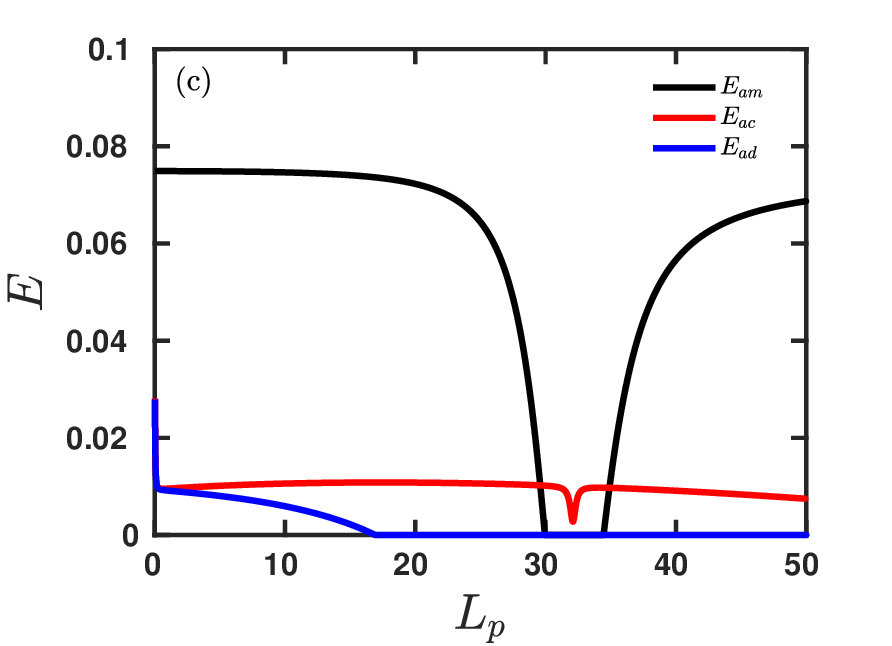}
	\caption{(Color online)\label{fig:entanglement_Delta_l} The bipartite entanglement between the optical and the mirror mode is plotted as a function of (a) cavity detuning when the orbital angular momentum of the input field $l=243$ (b) the OAM of the driving field for $L_p=1~\textrm{and}$ (c) the winding number of BEC for $l=243$ while the cavity detuning is $~\Delta'=-1.2\omega_\phi$. We consider the resonance frequency of the rotating mirror to be  $\omega_\phi=3$ MHz, mass of the mirror $M=0.1$ ng, radius $R_m=20$ $\mu$m, the cavity length $L=1$ mm, $\gamma_0/2\pi=0.48$ MHz, $\gamma_m/2\pi=0.8$ Hz, $\gamma_\phi/2\pi=4.77$ Hz, $\omega_0/2\pi=10^{15}$ Hz, $G/2\pi=7.67$ kHz, $U_0/2\pi=153.5$ Hz, $\Delta_a=1.04$ GHz, $\omega_\rho/2\pi=8.4$ kHz, and $a_{Na}=2.5$ nm. The input power of the driving field is $0.19$ nW. The bath temperature of the rotating mirror is $T_\phi = 5$ mK and the temperature of the atomic side modes are $T=10$ nK. }
\end{figure}  

In Fig. \ref{fig:entanglement_Delta_l}(a), we study the influence of the cavity detuning on the bipartite entanglements in our hybrid system. $E_{am}$, $E_{ac}$, and $E_{ad}$ denote the bipartite entanglements between cavity-mirror, cavity-atomic c mode, and cavity-atomic d mode, respectively. The black solid curve suggests the optimal cavity-mirror entanglement occurs around $\Delta'\approx-0.6\omega_\phi$. However, the blue and red solid curves represent the optimal values of $E_{ac}$ and $E_{ad}$ at $\Delta'\approx-1.2\omega_\phi$. The entanglement between the cavity and the atomic side modes sustains up to the bath temperature of the rotating mirror $T_\phi\approx13$ mK, as presented by the inset of Fig. \ref{fig:entanglement_Delta_l}(a).
In Fig. \ref{fig:entanglement_Delta_l}(b) and (c), we investigate the influence of the topological charge of the input optical drive and atomic rotation on the entanglement properties between various bipartite subsystems, respectively. The prominent interaction strength between the optical and the acoustic modes produces a more significant entanglement response than cavity-atomic side modes. More interestingly, our study predicts diminishing entanglements at a specific region of the OAM of the input beam. Also, the black solid curve of Fig. \ref{fig:entanglement_Delta_l}(c) suggests for the topological charge of the input beam $l=243$ {\color{blue}\cite{Zhou:24,10.1063/5.0089735, Ni:22, Pinnell_2021,He2022}}, the entanglement between the optical and the acoustic mode diminishes when the $L_p$ values lie between 30 and 34. Moreover, the two Bragg scattered atomic side modes produce distinct entanglement responses arising from the optimal cavity detuning condition $\Delta'=-1.2\omega_\phi=-\Omega_{c,d}$. Now, to explain the diminishing entanglements, we determine the energy of the rotating mirror, which is given by
\begin{align}
\label{eq:energy}
U&=\frac{\hbar \omega_\phi}{2}\left[\langle\delta\phi^2\rangle + \langle\delta L_z^2\rangle\right]=\frac{\hbar\omega_\phi}{2}\left(V_{77}+V_{88}\right)\nonumber\\
&=\hbar\omega_\phi\left(n_{eff}+\frac{1}{2}\right),
\end{align}
where the steady-state mean phonon number ($n_{eff}$) is associated with the effective mirror temperature ($T_{eff}$) by the relation $n_{eff}=\left(\textrm{exp}({\hbar\omega_\phi/k_B T_{eff}})-1\right)^{-1}$. 
\begin{figure}[h!]
	\includegraphics[width=0.46\textwidth]{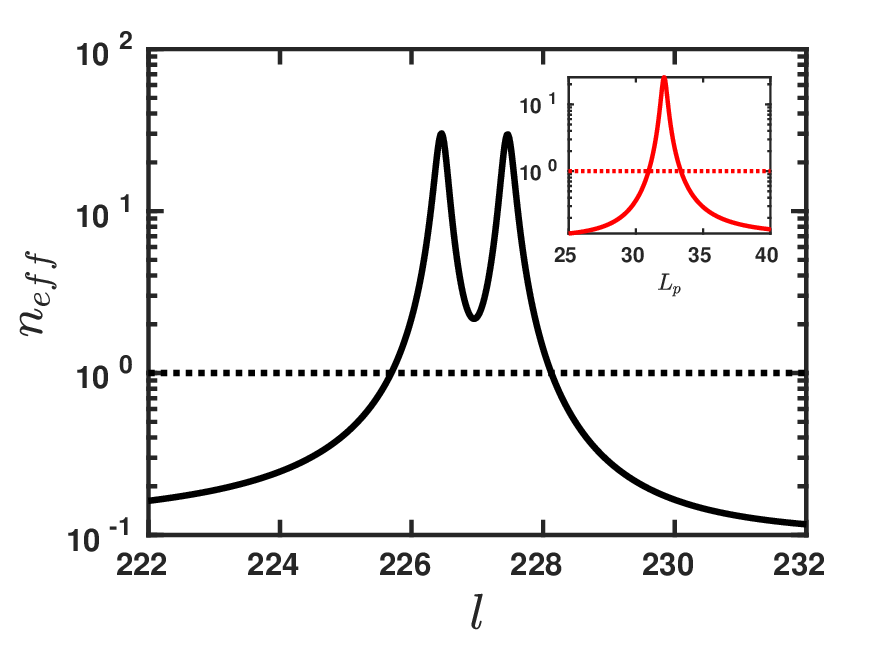}
	\label{fig8}
	\caption{\label{fig:phonon}(Color online) The mean phonon number of the rotating mirror as a function of the topological charge of the input field when $L_p=1$. The inset presents the variation of effective phonon number as a function of the winding number of BEC for $l=243$. The dotted lines correspond to the unity of the effective phonon number. All the other parameters are the same as in Fig. \ref{fig:entanglement_Delta_l}.  }
\end{figure}      
In Fig. \ref{fig:phonon} and in its inset, we present the effect of the OAM of the driving field and the angular momentum of atomic BEC on the steady-state phonon number, respectively. The solid black curve anticipates two distinct peaks in the effective phonon response stemming from the cooling inhibition associated with the topological charge $l\approx226$ and $227$, and the red solid curve of the inset depicts the suppression of cooling when the winding number $L_p$ lies between 30 and 34. In the subsequent discussion, we demonstrate how the occurrence of the dark modes suppresses the cooling mechanism by introducing a center of mass coordinate $(X_{1cm}, P_{1cm})$ and a relative coordinate $(X_{1r}, P_{1r})$ as
\begin{align}
\label{eq:centre_of_mass}
X_{1cm}=\frac{G X_d+g_\phi \phi}{\sqrt{G^2+g_\phi^2}}, P_{1cm}=\frac{G Y_d+g_\phi L_z}{\sqrt{G^2+g_\phi^2}},\nonumber\\
X_{1r}=\frac{G \phi-g_\phi X_d}{\sqrt{G^2+g_\phi^2}}, P_{1r}=\frac{G L_z-g_\phi Y_d}{\sqrt{G^2+g_\phi^2}}.
\end{align} 
Neglecting the atom-atom interaction, we can further express the Hamiltonian of Eq. \ref{eq6} as
\begin{widetext}
\begin{align}
\label{hamiltonian_relative}
H&=-\hbar\left(\Delta_0-\frac{U_0N}{2}\right)a^\dagger a - i\hbar\eta(a-a^\dagger)+\frac{\hbar\omega_c}{2}\left(X_c^2+Y_c^2\right)+\hbar G X_c a^\dagger a+\frac{\hbar}{2}\left(\frac{\omega_dG^2+\omega_\phi g_\phi^2}{G^2+g_\phi^2}\right)\left(X_{1cm}^2+P_{1cm}^2\right)\nonumber\\
&+\hbar\sqrt{G^2+g_\phi^2}X_{1cm}a^\dagger a+\frac{\hbar}{2}\left(\frac{\omega_dg_\phi^2+\omega_\phi G^2}{G^2+g_\phi^2}\right)\left(X_{1r}^2+P_{1r}^2\right)
+\frac{\hbar}{2} \frac{G g_\phi}{G^2+g_\phi^2}(\omega_\phi-\omega_d)(X_{1cm}X_{1r}+X_{1r}X_{1cm}+P_{1cm}P_{1r}\nonumber\\
&+P_{1r}P_{1cm}),
\end{align}
\end{widetext}
where the sixth and the last terms correspond to the interaction of the center of mass coordinates with the optical field and the relative coordinate, respectively. The above analysis shows that when $L_p=1~\textrm{and}~l\approx226$, one of the atomic side mode frequency $\omega_d$ matches with $\omega_\phi$ and the relative coordinate is decoupled from the center of mass coordinate as well as the optical mode. Nonetheless, it is straightforward to show the existence of another set of center of mass and relative coordinates defined as
\begin{align}
\label{eq:centre_of_mass2}
X_{2cm}=\frac{G X_c+g_\phi \phi}{\sqrt{G^2+g_\phi^2}}, P_{2cm}=\frac{G Y_c+g_\phi L_z}{\sqrt{G^2+g_\phi^2}},\nonumber\\
X_{2r}=\frac{G \phi-g_\phi X_c}{\sqrt{G^2+g_\phi^2}}, P_{2r}=\frac{G L_z-g_\phi Y_c}{\sqrt{G^2+g_\phi^2}},
\end{align} 
such that when $\omega_c=\omega_\phi$, the relative coordinate decouples from the center of mass and the cavity field. Hence, the radiation torque cooling is suppressed when the two atomic side modes degenerate with the acoustic mode ($\omega_\phi=\omega_{c,d}$) as the relative coordinate stays in the initial thermal state. Comparing Fig. \ref{fig:entanglement_Delta_l}(b) and Fig. \ref{fig:phonon}, we can say that cooling a rotating mirror close to its quantum ground state (\( l < 225 \) and \( l > 230 \)) helps to maintain quantum correlations. As a result, entanglement persists. However, between $l=226$ and $227$, the phonon number attains a minimum value of $2.2$, which signifies the proximity of the quantum ground state. Even at these low phonon numbers, the quantum fluctuations are sufficiently strong to disrupt the coherence necessary to attain any entanglement.
\subsection{Tripartite Entanglement}
Lastly, we study the tripartite entanglements between the cavity, mirror, and atomic side modes by quantifying the minimum residual contangle as \cite{Adesso_2007, Adesso_2006}
\begin{align}
\mathcal R_\tau^{\textrm{min}}&=\textrm{min}\left[\mathcal{R}_\tau^{a|mk},\mathcal{R}_\tau^{m|ak},\mathcal{R}_\tau^{k|am}\right]\;,
\end{align}
where, $\mathcal{R}_\tau^{i|jk}=C_{i|jk}-C_{i|j}-C_{i|k}\ge 0 ~~(i,j,k=\textrm{cavity}(a),\textrm{mirror}(m), ~\textrm{atomic}~c~\textrm{or}~d~\textrm{modes})$, is the residual {\color{black}contangle} written in terms of contangle  $C_{u|v}$ of subsystems of $u~\textrm{and}~v$ (see Ref. \cite{PhysRevLett.121.203601} for more details of calculating $\mathcal{R}_\tau^\textrm{min}$). The black solid and the red solid curve of Fig. \ref{fig:tripartite}(a) suggest that the tripartite entanglements between the three constituents of the model system (cavity, mirror, and c-or d-modes) are optimized when the cavity detuning $\Delta'=-1.9\omega_\phi$.  
\begin{figure}[h!]
	\includegraphics[width=0.46\textwidth]{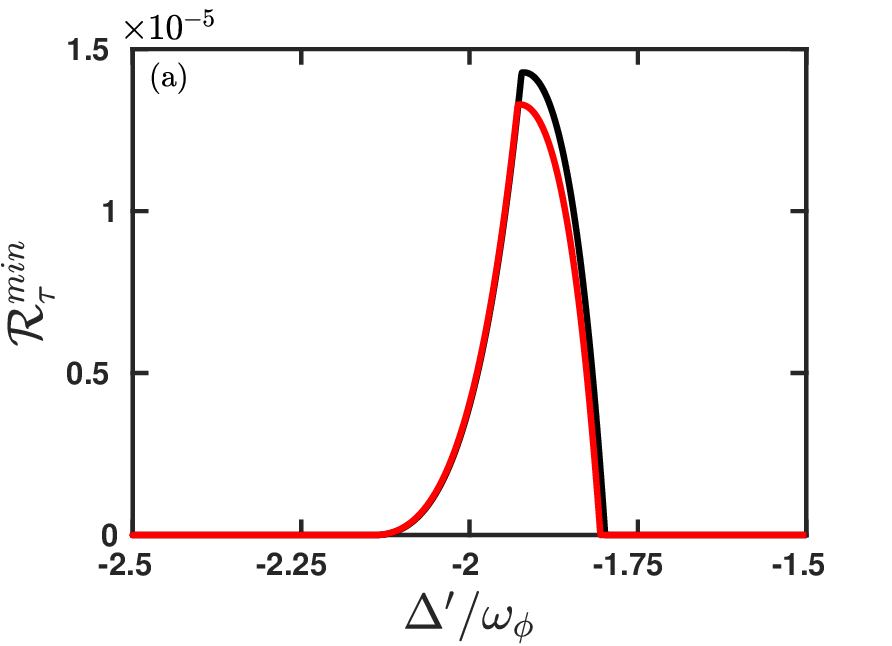}
	\includegraphics[width=0.46\textwidth]{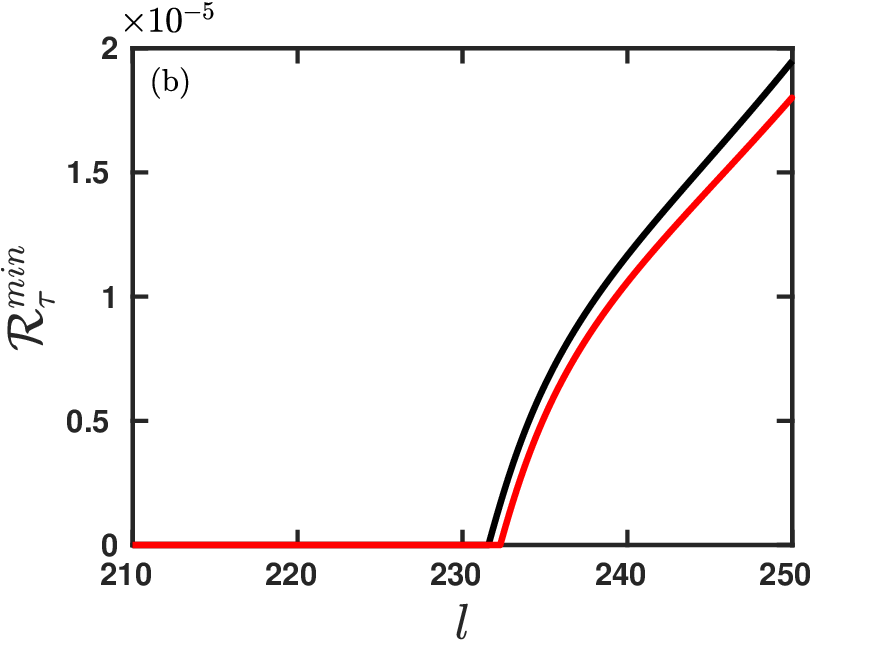}
    \includegraphics[width=0.46\textwidth]{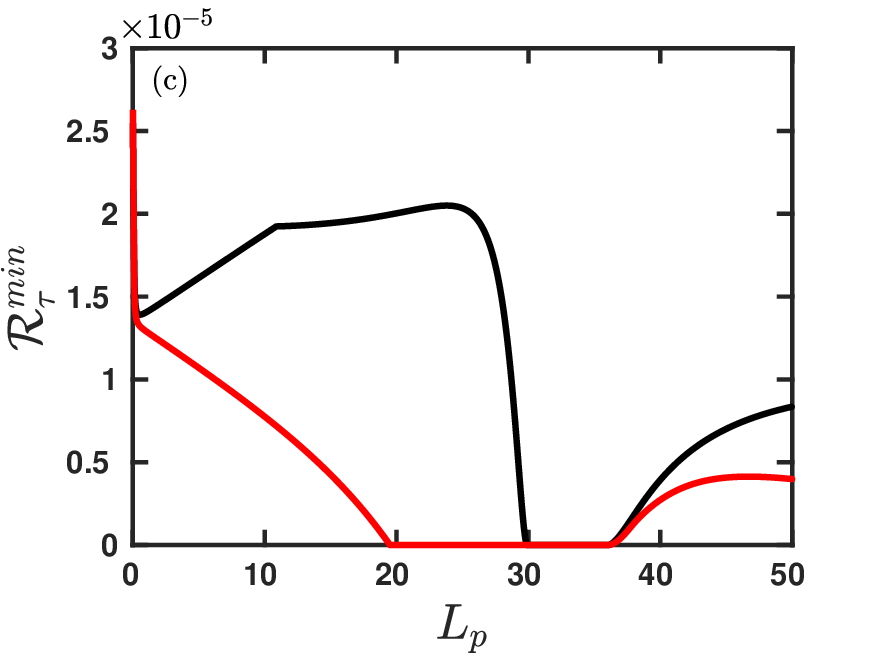}
	\caption{(Color online)\label{fig:tripartite} Tripartite entanglement in terms of minimum residual contangle versus (a) cavity detuning for $l=243$, (b) the OAM of the driving field for $L_p=1~\textrm{and}~\Delta'=-1.9\omega_\phi$. The black solid curve accounts for the tripartite entanglement associated with the cavity-mirror-atomic c mode. The red solid curve accounts for the tripartite entanglement associated with the cavity-mirror-atomic d mode. All the other parameters are same as in Fig. \ref{fig:entanglement_Delta_l}.}
\end{figure}  
Moreover, the tripartite entanglement arises when there's frequent optical interaction with matter waves, particularly in proportion to the number of lattice maxima $(2l)$. It means that the likelihood of optical interaction increases as the number of lattice maxima increases. When such interactions become frequent, they can lead to the emergence of tripartite entanglement, as shown in Fig. \ref{fig:tripartite}(b). In Fig. \ref{fig:tripartite}(c), we present the tripartite entanglement between the cavity, mirror, and atomic side modes as a function of the winding number of the condensate atoms when the topological charge of the driving field is $l=243$. The black solid curve presents the diminishing effect of the three body entanglement around $L_p=30$ due to the thermal fluctuations of the rotating mirror. The red solid curve depicts the tripartite entanglement response between the cavity, mirror, and the atomic d-mode. The presence of the atomic d-mode produces a wider range in $L_p$, where the tripartite entanglement does not persist. Moreover, the very distinct three-body entanglement response can be utilized to distinguish clearly between the two atomic side modes.

    \section{Conclusion}
        \label{sec:conc}
     We have presented an unique hybrid opto-rotational system consisting of a ring BEC confined inside a LG cavity. The system shows a distinctive way to squeeze the quantum fluctuations of the output light field quadratures below the shot noise. With the experimentally feasible parameters and for a measurement angle of $7^\circ$, we achieved 87$\%$ of the ponderomotive squeezing around the frequencies of Bragg-scattered sidemodes.  Furthermore, two very distinct systems in our hybrid configuration get coupled to a common cavity mode. As a consequence, optical squeezing of about $40\%$ occurs even at the rotating mirror frequency and can further be manipulated by the persistent currents in the annularly trapped BEC. Our scheme also provides a versatile pathway to utilize the atomic interactions and the radiation pressure force to produce bipartite and tripartite entanglement between different physical entities of the hybrid system. Interestingly, the quantum correlations in the system last under the cooling conditions of the rotating end mirror and correspondingly there exists a parameter regime where the entanglement survives. We expect our results find interesting applications in atomtronics \cite{Amico2021}, optomechanics \cite{Kippenberg_OptExp2007}, sensing \cite{Li_NanoPhoto2021} and quantum information processing \cite{Stannigel_PRL2012}.

    \section*{Acknowledgments}
    We thank R. Kanamoto for discussions. P. K. acknowledges the financial support from the Max Planck Society. M.B. thanks the Air Force Office of Scientific Research (FA9550-23-1-0259) for support. 
    
     \appendix
\section{Power Spectral Density}
\label{AppendixA}
{\color{black}
To obtain the output quadrature noise spectrum, we use following input-output relation,
\begin{align}
\label{eq:output}
a_{out}(\omega)&=\sqrt{\gamma_{o}}\delta a(\omega)-\delta a_{in}(\omega)\;.
\end{align}
From Eq. (\ref{eq:output}), we can write following quadrature relations
\begin{align}
\mathcal{Q}_{out}(\omega)&=\sqrt{\gamma_{o}}\delta \mathcal{Q}(\omega)-\mathcal{Q}_{in}(\omega)\;,\label{eqA2}\\
\mathcal{P}_{out}(\omega)&=\sqrt{\gamma_{o}}\delta \mathcal{P}(\omega)-\mathcal{P}_{in}(\omega)\;,\label{eqA3}
\end{align}
By Fourier transforming the Eq. (\ref{eq:fluctuation_eq}), we obtain the output optical quadratures as 
\begin{widetext}
\begin{align}
\label{eq:output_quadrature}
\mathcal{Q}_{out}(\omega)&=\Big[\sqrt{\gamma_{o}}\tilde{F}_{2}(\omega)-1\Big]\mathcal{Q}_{in}(\omega)+i\sqrt{\gamma_{o}}\tilde{F}_{3}(\omega)\mathcal{P}_{in}(\omega)+\sqrt{2\gamma_{o}}\tilde{F}_{5}(\omega)\epsilon_{c}(\omega)+\sqrt{2\gamma_{o}}\tilde{F}_{7}(\omega)\epsilon_{d}(\omega)+\sqrt{2\gamma_{o}}\tilde{F}_{9}(\omega)\epsilon_{\phi}(\omega)\;,\nonumber\\
\mathcal{P}_{out}(\omega)&=-i\sqrt{\gamma_{o}}\Big[2\tilde{F}_{1}(\omega)+\tilde{F}_{3}(\omega)\Big]\mathcal{Q}_{in}(\omega)+\Big[\sqrt{\gamma_{o}}\tilde{F}_{2}(\omega)-1\Big]\mathcal{P}_{in}(\omega)-i\sqrt{2\gamma_{o}}\tilde{F}_{4}(\omega)\epsilon_{c}(\omega)-i\sqrt{2\gamma_{o}}\tilde{F}_{6}(\omega)\epsilon_{d}(\omega)\nonumber\\
&-i\sqrt{2\gamma_{o}}\tilde{F}_{8}(\omega)\epsilon_{\phi}(\omega)\;,
\end{align}
\end{widetext}
where $\tilde F_i(\omega)$'s are complicated complex functions written as below.
\begin{widetext}
\begin{align}
\label{eq:F_i}
\tilde{F_1}(\omega)&=\frac{i a_s^2\sqrt{\gamma_0}}{D(\omega)}\left[G^2A_\phi(\omega)\left[\tilde{\omega}_c\left(\mathcal{A}+\tilde{A_d}(\omega)\right)+\tilde{\omega}_d\left(\tilde{A}_c(\omega)-\mathcal{A}\right)\right]+\omega_\phi g_\phi^2\left[\mathcal{A}^2+\tilde{A}_c(\omega)\tilde{A}_d(\omega)\right]\right],\nonumber\\
\tilde{F_2}(\omega)&=\frac{\sqrt{\gamma_0}}{D(\omega)}\left[\mathcal{A}^2+\tilde{A}_c(\omega)\tilde{A}_d(\omega)\right]A_3(\omega)A_\phi(\omega),
~\tilde{F_3}(\omega)=\frac{i\Delta'\sqrt{\gamma_0}}{D(\omega)}A_\phi(\omega)\left[\mathcal{A}^2+\tilde{A}_c(\omega)\tilde{A}_d(\omega)\right],\nonumber\\
\tilde{F_4}(\omega)&=-\frac{iGa_s\Omega_c}{D(\omega)}A_3(\omega)A_\phi(\omega)\left[\mathcal{A}+\tilde{A}_d(\omega)\right],
~\tilde{F_5}(\omega)=\frac{\Delta'\Omega_c G a_s}{D(\omega)}A_\phi(\omega)\left[\mathcal{A}+\tilde{A}_d(\omega)\right],\nonumber\\
\tilde{F_6}(\omega)&=-\frac{iGa_s\Omega_d}{D(\omega)}A_3(\omega)A_\phi(\omega)\left[\tilde{A}_c(\omega)-\mathcal{A}\right],
~\tilde{F_7}(\omega)=\frac{\Delta'\Omega_d G a_s}{D(\omega)}A_\phi(\omega)\left[-\mathcal{A}+\tilde{A}_c(\omega)\right],\nonumber\\
\tilde{F_8}(\omega)&=-\frac{ig_\phi a_s\omega_\phi}{D(\omega)}A_3(\omega)\left[\mathcal{A}^2+\tilde{A}_c(\omega)\tilde{A}_d(\omega)\right],
~\tilde{F_9}(\omega)=\frac{\Delta'\omega_\phi g_\phi a_s}{D(\omega)}\left[\mathcal{A}^2+\tilde{A}_c(\omega)\tilde{A}_d(\omega)\right].
\end{align}
\end{widetext}
Further, using Eq. (\ref{eq:output_quadrature}) in Eq. (\ref{eq:Homodyne}), we obtain
\begin{align}
\label{eq:Homodyne_amplitude}
\mathcal{Q}_{\theta}^{out}(\omega)&=\xi_{1}(\omega)\mathcal{Q}_{in}(\omega)+\xi_{2}(\omega)\mathcal{P}_{in}(\omega)
+\xi_{3}(\omega)\epsilon_{c}(\omega)\nonumber\\
&+\xi_{4}(\omega)\epsilon_{d}(\omega)+\xi_{5}(\omega)\epsilon_{\phi}(\omega)\;,
\end{align}
where, $\xi_i$'s can be expressed as
\begin{align}
\label{eq:defininition_xi}
\xi_{1}(\omega)&=-i\sqrt{\gamma_{o}}\left[2\tilde{F}_{1}(\omega)+\tilde{F}_{3}(\omega)\right]\sin\theta\nonumber\\
&+\left[\sqrt{\gamma_{o}}\tilde{F}_{2}(\omega)-1\right]\cos\theta\;,\nonumber\\
\xi_{2}(\omega)&=\left[\sqrt{\gamma_{o}}\tilde{F}_{2}(\omega)-1\right]\sin\theta+i\sqrt{\gamma_{o}}\tilde{F}_{3}(\omega)\cos\theta\;,\nonumber\\
\xi_{3}(\omega)&=-i\sqrt{2\gamma_{o}}\tilde{F}_{4}(\omega)\sin\theta+\sqrt{2\gamma_{o}}\tilde{F}_{5}(\omega)\cos\theta\;,\nonumber\\
\xi_{4}(\omega)&=-i\sqrt{2\gamma_{o}}\tilde{F}_{6}(\omega)\sin\theta+\sqrt{2\gamma_{o}}\tilde{F}_{7}(\omega)\cos\theta\;,\nonumber\\
\xi_{5}(\omega)&=-i\sqrt{2\gamma_{o}}\tilde{F}_{8}(\omega)\sin\theta+\sqrt{2\gamma_{o}}\tilde{F}_{9}(\omega)\cos\theta\;.
\end{align}
 The output quadrature spectrum is obtained from the definition 
 \begin{align}
 \label{eq:psd}
       S(\omega)&=\frac{1}{\pi}\int_{-\infty}^{\infty}d\omega'\langle \mathcal{Q}_\theta^{out}(\omega')\mathcal{Q}_\theta^{out}(\omega)\rangle\;.\nonumber\\
 \end{align}
By using the following correlation relations
\begin{align}
\label{eq:correlation_relations}
\langle \mathcal{Q}_{in}(\omega^{'})\mathcal{Q}_{in}(\omega)\rangle&=\langle \mathcal{P}_{in}(\omega^{'})\mathcal{P}_{in}(\omega)\rangle=\pi\delta(\omega^{'}+\omega)\;,\nonumber\\
\langle \mathcal{Q}_{in}(\omega^{'})\mathcal{P}_{in}(\omega)\rangle&=-\langle \mathcal{P}_{in}(\omega^{'})\mathcal{Q}_{in}(\omega)\rangle=i\pi\delta(\omega^{'}+\omega)\;,\nonumber\\
\langle \epsilon_{c,d}(\omega^{'})\epsilon_{c,d}(\omega)\rangle&=2\pi\frac{\gamma_{m}\omega^{'}}{\Omega_{c,d}}\Bigg[1+\coth\Big(\frac{\hbar\omega^{'}}{2k_{B}T}\Big)\Bigg]\delta(\omega^{'}+\omega)\;,\nonumber\\
\langle \epsilon_{\phi}(\omega^{'})\epsilon_{\phi}(\omega)\rangle&=2\pi\frac{\gamma_{\phi}\omega^{'}}{\omega_{\phi}}\Bigg[1+\coth\Big(\frac{\hbar\omega^{'}}{2k_{B}T_{\phi}}\Big)\Bigg]\delta(\omega^{'}+\omega)\;,
\end{align}
we obtain the analytical expression of Eq. \ref{eq:psd_definition}.
\vspace{-1mm}
\section{Optimized Squeezing}
\label{AppendixB}
The optimized angle in Eq. (\ref{eq:optimized_theta}) is expressed in terms of $B_{1}(\omega)$ and $B_{2}(\omega)$ and their expressions are given below
\begin{widetext}
\begin{align}
B_1(\omega)&=|\kappa_1(\omega)|^2-|\kappa_3(\omega)|^2+ i\left(\kappa_1^*(\omega)\kappa_2(\omega)-\kappa_1(\omega)\kappa_2^*(\omega)-\kappa_2^*(\omega)\kappa_3(\omega)+\kappa_3^*(\omega)\kappa_2(\omega)\right)\nonumber\\
&-2\gamma_m\frac{\omega}{\Omega_c}\left[1-\coth\left(\frac{\hbar\omega}{2k_B T}\right)\right]\left(|\kappa_4(\omega)|^2-|\kappa_5(\omega)|^2\right)
-2\gamma_m\frac{\omega}{\Omega_d}\left[1-\coth\left(\frac{\hbar\omega}{2k_B T}\right)\right]\left(|\kappa_6(\omega)|^2-|\kappa_7(\omega)|^2\right)\nonumber\\
&-2\gamma_\phi\frac{\omega}{\omega_\phi}\left[1-\coth\left(\frac{\hbar\omega}{2k_B T_\phi}\right)\right]\left(|\kappa_8(\omega)|^2-|\kappa_9(\omega)|^2\right),\nonumber\\
\\
B_2(\omega)=&\kappa_1^*(\omega)\kappa_2(\omega)+\kappa_2^*(\omega)\kappa_1(\omega)+\kappa_2^*(\omega)\kappa_3(\omega)+\kappa_2(\omega)\kappa_3^*(\omega)+i\left(\kappa_1^*(\omega)\kappa_3(\omega)-\kappa_3^*(\omega)\kappa_1(\omega)\right)\nonumber\\
&-2\gamma_m\frac{\omega}{\Omega_c}\left[1-\coth\left(\frac{\hbar\omega}{2k_B T}\right)\right]\left(\kappa_4^*(\omega)\kappa_5(\omega)+\kappa_4(\omega)\kappa_5^*(\omega)\right)\nonumber\\
&-2\gamma_m\frac{\omega}{\Omega_d}\left[1-\coth\left(\frac{\hbar\omega}{2k_B T}\right)\right]\left(\kappa_6^*(\omega)\kappa_7(\omega)+\kappa_6(\omega)\kappa_7^*(\omega)\right)\nonumber\\
&-2\gamma_\phi\frac{\omega}{\omega_\phi}\left[1-\coth\left(\frac{\hbar\omega}{2k_BT_\phi}\right)\right]\left(\kappa_9(\omega)\kappa_8^*(\omega)+\kappa_9^*(\omega)\kappa_8(\omega)\right)
 \end{align}
where $\kappa_{i}$'s are written as 
\begin{align}
\kappa_{1}(\omega)=-i\sqrt{\gamma_{o}}\Big[2\tilde{F}_{1}(\omega)+\tilde{F}_{3}(\omega)\Big]\;,
\kappa_{2}(\omega)=\sqrt{\gamma_{o}}\tilde{F}_{2}(\omega)-1\;,
\kappa_{3}(\omega)=i\sqrt{\gamma_{o}}\tilde{F}_{3}(\omega)\;,\nonumber\\
\kappa_{4}(\omega)=-i\sqrt{2\gamma_{o}}\tilde{F}_{4}(\omega)\;,
\kappa_{5}(\omega)=\sqrt{2\gamma_{o}}\tilde{F}_{5}(\omega)\;,
\kappa_{6}(\omega)=-i\sqrt{2\gamma_{o}}\tilde{F}_{6}(\omega)\;,\nonumber\\
\kappa_{7}(\omega)=\sqrt{2\gamma_{o}}\tilde{F}_{7}(\omega)\;,
\kappa_{8}(\omega)=-i\sqrt{2\gamma_{o}}\tilde{F}_{8}(\omega)\;,
\kappa_{9}(\omega)=\sqrt{2\gamma_{o}}\tilde{F}_{9}(\omega)\;.  
\end{align}
\end{widetext}
        \nocite{*}

\begin{thebibliography}{79}%
	\makeatletter
	\providecommand \@ifxundefined [1]{%
		\@ifx{#1\undefined}
	}%
	\providecommand \@ifnum [1]{%
		\ifnum #1\expandafter \@firstoftwo
		\else \expandafter \@secondoftwo
		\fi
	}%
	\providecommand \@ifx [1]{%
		\ifx #1\expandafter \@firstoftwo
		\else \expandafter \@secondoftwo
		\fi
	}%
	\providecommand \natexlab [1]{#1}%
	\providecommand \enquote  [1]{``#1''}%
	\providecommand \bibnamefont  [1]{#1}%
	\providecommand \bibfnamefont [1]{#1}%
	\providecommand \citenamefont [1]{#1}%
	\providecommand \href@noop [0]{\@secondoftwo}%
	\providecommand \href [0]{\begingroup \@sanitize@url \@href}%
	\providecommand \@href[1]{\@@startlink{#1}\@@href}%
	\providecommand \@@href[1]{\endgroup#1\@@endlink}%
	\providecommand \@sanitize@url [0]{\catcode `\\12\catcode `\$12\catcode `\&12\catcode `\#12\catcode `\^12\catcode `\_12\catcode `\%12\relax}%
	\providecommand \@@startlink[1]{}%
	\providecommand \@@endlink[0]{}%
	\providecommand \url  [0]{\begingroup\@sanitize@url \@url }%
	\providecommand \@url [1]{\endgroup\@href {#1}{\urlprefix }}%
	\providecommand \urlprefix  [0]{URL }%
	\providecommand \Eprint [0]{\href }%
	\providecommand \doibase [0]{https://doi.org/}%
	\providecommand \selectlanguage [0]{\@gobble}%
	\providecommand \bibinfo  [0]{\@secondoftwo}%
	\providecommand \bibfield  [0]{\@secondoftwo}%
	\providecommand \translation [1]{[#1]}%
	\providecommand \BibitemOpen [0]{}%
	\providecommand \bibitemStop [0]{}%
	\providecommand \bibitemNoStop [0]{.\EOS\space}%
	\providecommand \EOS [0]{\spacefactor3000\relax}%
	\providecommand \BibitemShut  [1]{\csname bibitem#1\endcsname}%
	\let\auto@bib@innerbib\@empty
	\bibitem [{\citenamefont {Anderson}\ \emph {et~al.}(1995)\citenamefont {Anderson}, \citenamefont {Ensher}, \citenamefont {Matthews}, \citenamefont {Wieman},\ and\ \citenamefont {Cornell}}]{doi:10.1126/science.269.5221.198}%
	\BibitemOpen
	\bibfield  {author} {\bibinfo {author} {\bibfnamefont {M.~H.}\ \bibnamefont {Anderson}}, \bibinfo {author} {\bibfnamefont {J.~R.}\ \bibnamefont {Ensher}}, \bibinfo {author} {\bibfnamefont {M.~R.}\ \bibnamefont {Matthews}}, \bibinfo {author} {\bibfnamefont {C.~E.}\ \bibnamefont {Wieman}},\ and\ \bibinfo {author} {\bibfnamefont {E.~A.}\ \bibnamefont {Cornell}},\ }\bibfield  {title} {\bibinfo {title} {Observation of bose-einstein condensation in a dilute atomic vapor},\ }\href {https://doi.org/10.1126/science.269.5221.198} {\bibfield  {journal} {\bibinfo  {journal} {Science}\ }\textbf {\bibinfo {volume} {269}},\ \bibinfo {pages} {198} (\bibinfo {year} {1995})},\ \Eprint {https://arxiv.org/abs/https://www.science.org/doi/pdf/10.1126/science.269.5221.198} {https://www.science.org/doi/pdf/10.1126/science.269.5221.198} \BibitemShut {NoStop}%
	\bibitem [{\citenamefont {Davis}\ \emph {et~al.}(1995)\citenamefont {Davis}, \citenamefont {Mewes}, \citenamefont {Andrews}, \citenamefont {van Druten}, \citenamefont {Durfee}, \citenamefont {Kurn},\ and\ \citenamefont {Ketterle}}]{davis1995bose}%
	\BibitemOpen
	\bibfield  {author} {\bibinfo {author} {\bibfnamefont {K.~B.}\ \bibnamefont {Davis}}, \bibinfo {author} {\bibfnamefont {M.-O.}\ \bibnamefont {Mewes}}, \bibinfo {author} {\bibfnamefont {M.~R.}\ \bibnamefont {Andrews}}, \bibinfo {author} {\bibfnamefont {N.~J.}\ \bibnamefont {van Druten}}, \bibinfo {author} {\bibfnamefont {D.~S.}\ \bibnamefont {Durfee}}, \bibinfo {author} {\bibfnamefont {D.}~\bibnamefont {Kurn}},\ and\ \bibinfo {author} {\bibfnamefont {W.}~\bibnamefont {Ketterle}},\ }\bibfield  {title} {\bibinfo {title} {Bose-einstein condensation in a gas of sodium atoms},\ }\href@noop {} {\bibfield  {journal} {\bibinfo  {journal} {Physical review letters}\ }\textbf {\bibinfo {volume} {75}},\ \bibinfo {pages} {3969} (\bibinfo {year} {1995})}\BibitemShut {NoStop}%
	\bibitem [{\citenamefont {Leggett}(2001)}]{RevModPhys.73.307}%
	\BibitemOpen
	\bibfield  {author} {\bibinfo {author} {\bibfnamefont {A.~J.}\ \bibnamefont {Leggett}},\ }\bibfield  {title} {\bibinfo {title} {Bose-einstein condensation in the alkali gases: Some fundamental concepts},\ }\href {https://doi.org/10.1103/RevModPhys.73.307} {\bibfield  {journal} {\bibinfo  {journal} {Rev. Mod. Phys.}\ }\textbf {\bibinfo {volume} {73}},\ \bibinfo {pages} {307} (\bibinfo {year} {2001})}\BibitemShut {NoStop}%
	\bibitem [{\citenamefont {Vilchynskyy}\ \emph {et~al.}(2013)\citenamefont {Vilchynskyy}, \citenamefont {Yakimenko}, \citenamefont {Isaieva},\ and\ \citenamefont {Chumachenko}}]{vilchynskyy2013nature}%
	\BibitemOpen
	\bibfield  {author} {\bibinfo {author} {\bibfnamefont {S.}~\bibnamefont {Vilchynskyy}}, \bibinfo {author} {\bibfnamefont {A.}~\bibnamefont {Yakimenko}}, \bibinfo {author} {\bibfnamefont {K.}~\bibnamefont {Isaieva}},\ and\ \bibinfo {author} {\bibfnamefont {A.}~\bibnamefont {Chumachenko}},\ }\bibfield  {title} {\bibinfo {title} {The nature of superfluidity and bose-einstein condensation: From liquid 4he to dilute ultracold atomic gases},\ }\href@noop {} {\bibfield  {journal} {\bibinfo  {journal} {Low Temperature Physics}\ }\textbf {\bibinfo {volume} {39}},\ \bibinfo {pages} {724} (\bibinfo {year} {2013})}\BibitemShut {NoStop}%
	\bibitem [{\citenamefont {Hashimoto}\ \emph {et~al.}(2020)\citenamefont {Hashimoto}, \citenamefont {Ota}, \citenamefont {Tsuzuki}, \citenamefont {Nagashima}, \citenamefont {Fukushima}, \citenamefont {Kasahara}, \citenamefont {Matsuda}, \citenamefont {Matsuura}, \citenamefont {Mizukami}, \citenamefont {Shibauchi}, \citenamefont {Shin},\ and\ \citenamefont {Okazaki}}]{doi:10.1126/sciadv.abb9052}%
	\BibitemOpen
	\bibfield  {author} {\bibinfo {author} {\bibfnamefont {T.}~\bibnamefont {Hashimoto}}, \bibinfo {author} {\bibfnamefont {Y.}~\bibnamefont {Ota}}, \bibinfo {author} {\bibfnamefont {A.}~\bibnamefont {Tsuzuki}}, \bibinfo {author} {\bibfnamefont {T.}~\bibnamefont {Nagashima}}, \bibinfo {author} {\bibfnamefont {A.}~\bibnamefont {Fukushima}}, \bibinfo {author} {\bibfnamefont {S.}~\bibnamefont {Kasahara}}, \bibinfo {author} {\bibfnamefont {Y.}~\bibnamefont {Matsuda}}, \bibinfo {author} {\bibfnamefont {K.}~\bibnamefont {Matsuura}}, \bibinfo {author} {\bibfnamefont {Y.}~\bibnamefont {Mizukami}}, \bibinfo {author} {\bibfnamefont {T.}~\bibnamefont {Shibauchi}}, \bibinfo {author} {\bibfnamefont {S.}~\bibnamefont {Shin}},\ and\ \bibinfo {author} {\bibfnamefont {K.}~\bibnamefont {Okazaki}},\ }\bibfield  {title} {\bibinfo {title} {Bose-einstein condensation superconductivity induced by disappearance of the nematic state},\ }\href {https://doi.org/10.1126/sciadv.abb9052} {\bibfield  {journal} {\bibinfo  {journal} {Science
				Advances}\ }\textbf {\bibinfo {volume} {6}},\ \bibinfo {pages} {eabb9052} (\bibinfo {year} {2020})},\ \Eprint {https://arxiv.org/abs/https://www.science.org/doi/pdf/10.1126/sciadv.abb9052} {https://www.science.org/doi/pdf/10.1126/sciadv.abb9052} \BibitemShut {NoStop}%
	\bibitem [{\citenamefont {Fetter}(2009)}]{RevModPhys.81.647}%
	\BibitemOpen
	\bibfield  {author} {\bibinfo {author} {\bibfnamefont {A.~L.}\ \bibnamefont {Fetter}},\ }\bibfield  {title} {\bibinfo {title} {Rotating trapped bose-einstein condensates},\ }\href {https://doi.org/10.1103/RevModPhys.81.647} {\bibfield  {journal} {\bibinfo  {journal} {Rev. Mod. Phys.}\ }\textbf {\bibinfo {volume} {81}},\ \bibinfo {pages} {647} (\bibinfo {year} {2009})}\BibitemShut {NoStop}%
	\bibitem [{\citenamefont {Mueller}\ \emph {et~al.}(1998)\citenamefont {Mueller}, \citenamefont {Goldbart},\ and\ \citenamefont {Lyanda-Geller}}]{PhysRevA.57.R1505}%
	\BibitemOpen
	\bibfield  {author} {\bibinfo {author} {\bibfnamefont {E.~J.}\ \bibnamefont {Mueller}}, \bibinfo {author} {\bibfnamefont {P.~M.}\ \bibnamefont {Goldbart}},\ and\ \bibinfo {author} {\bibfnamefont {Y.}~\bibnamefont {Lyanda-Geller}},\ }\bibfield  {title} {\bibinfo {title} {Multiply connected bose-einstein-condensed alkali-metal gases: Current-carrying states and their decay},\ }\href {https://doi.org/10.1103/PhysRevA.57.R1505} {\bibfield  {journal} {\bibinfo  {journal} {Phys. Rev. A}\ }\textbf {\bibinfo {volume} {57}},\ \bibinfo {pages} {R1505} (\bibinfo {year} {1998})}\BibitemShut {NoStop}%
	\bibitem [{\citenamefont {Das}\ \emph {et~al.}(2012)\citenamefont {Das}, \citenamefont {Sabbatini},\ and\ \citenamefont {Zurek}}]{Das2012}%
	\BibitemOpen
	\bibfield  {author} {\bibinfo {author} {\bibfnamefont {A.}~\bibnamefont {Das}}, \bibinfo {author} {\bibfnamefont {J.}~\bibnamefont {Sabbatini}},\ and\ \bibinfo {author} {\bibfnamefont {W.~H.}\ \bibnamefont {Zurek}},\ }\bibfield  {title} {\bibinfo {title} {Winding up superfluid in a torus via bose einstein condensation},\ }\href {https://doi.org/10.1038/srep00352} {\bibfield  {journal} {\bibinfo  {journal} {Scientific Reports}\ }\textbf {\bibinfo {volume} {2}},\ \bibinfo {pages} {352} (\bibinfo {year} {2012})}\BibitemShut {NoStop}%
	\bibitem [{\citenamefont {Beattie}\ \emph {et~al.}(2013)\citenamefont {Beattie}, \citenamefont {Moulder}, \citenamefont {Fletcher},\ and\ \citenamefont {Hadzibabic}}]{PhysRevLett.110.025301}%
	\BibitemOpen
	\bibfield  {author} {\bibinfo {author} {\bibfnamefont {S.}~\bibnamefont {Beattie}}, \bibinfo {author} {\bibfnamefont {S.}~\bibnamefont {Moulder}}, \bibinfo {author} {\bibfnamefont {R.~J.}\ \bibnamefont {Fletcher}},\ and\ \bibinfo {author} {\bibfnamefont {Z.}~\bibnamefont {Hadzibabic}},\ }\bibfield  {title} {\bibinfo {title} {Persistent currents in spinor condensates},\ }\href {https://doi.org/10.1103/PhysRevLett.110.025301} {\bibfield  {journal} {\bibinfo  {journal} {Phys. Rev. Lett.}\ }\textbf {\bibinfo {volume} {110}},\ \bibinfo {pages} {025301} (\bibinfo {year} {2013})}\BibitemShut {NoStop}%
	\bibitem [{\citenamefont {Guo}\ \emph {et~al.}(2020)\citenamefont {Guo}, \citenamefont {Dubessy}, \citenamefont {de~Herve}, \citenamefont {Kumar}, \citenamefont {Badr}, \citenamefont {Perrin}, \citenamefont {Longchambon},\ and\ \citenamefont {Perrin}}]{PhysRevLett.124.025301}%
	\BibitemOpen
	\bibfield  {author} {\bibinfo {author} {\bibfnamefont {Y.}~\bibnamefont {Guo}}, \bibinfo {author} {\bibfnamefont {R.}~\bibnamefont {Dubessy}}, \bibinfo {author} {\bibfnamefont {M.~d.~G.}\ \bibnamefont {de~Herve}}, \bibinfo {author} {\bibfnamefont {A.}~\bibnamefont {Kumar}}, \bibinfo {author} {\bibfnamefont {T.}~\bibnamefont {Badr}}, \bibinfo {author} {\bibfnamefont {A.}~\bibnamefont {Perrin}}, \bibinfo {author} {\bibfnamefont {L.}~\bibnamefont {Longchambon}},\ and\ \bibinfo {author} {\bibfnamefont {H.}~\bibnamefont {Perrin}},\ }\bibfield  {title} {\bibinfo {title} {Supersonic rotation of a superfluid: A long-lived dynamical ring},\ }\href {https://doi.org/10.1103/PhysRevLett.124.025301} {\bibfield  {journal} {\bibinfo  {journal} {Phys. Rev. Lett.}\ }\textbf {\bibinfo {volume} {124}},\ \bibinfo {pages} {025301} (\bibinfo {year} {2020})}\BibitemShut {NoStop}%
	\bibitem [{\citenamefont {Ryu}\ \emph {et~al.}(2007)\citenamefont {Ryu}, \citenamefont {Andersen}, \citenamefont {Clad\'e}, \citenamefont {Natarajan}, \citenamefont {Helmerson},\ and\ \citenamefont {Phillips}}]{PhysRevLett.99.260401}%
	\BibitemOpen
	\bibfield  {author} {\bibinfo {author} {\bibfnamefont {C.}~\bibnamefont {Ryu}}, \bibinfo {author} {\bibfnamefont {M.~F.}\ \bibnamefont {Andersen}}, \bibinfo {author} {\bibfnamefont {P.}~\bibnamefont {Clad\'e}}, \bibinfo {author} {\bibfnamefont {V.}~\bibnamefont {Natarajan}}, \bibinfo {author} {\bibfnamefont {K.}~\bibnamefont {Helmerson}},\ and\ \bibinfo {author} {\bibfnamefont {W.~D.}\ \bibnamefont {Phillips}},\ }\bibfield  {title} {\bibinfo {title} {Observation of persistent flow of a bose-einstein condensate in a toroidal trap},\ }\href {https://doi.org/10.1103/PhysRevLett.99.260401} {\bibfield  {journal} {\bibinfo  {journal} {Phys. Rev. Lett.}\ }\textbf {\bibinfo {volume} {99}},\ \bibinfo {pages} {260401} (\bibinfo {year} {2007})}\BibitemShut {NoStop}%
	\bibitem [{\citenamefont {Ramanathan}\ \emph {et~al.}(2011)\citenamefont {Ramanathan}, \citenamefont {Wright}, \citenamefont {Muniz}, \citenamefont {Zelan}, \citenamefont {Hill}, \citenamefont {Lobb}, \citenamefont {Helmerson}, \citenamefont {Phillips},\ and\ \citenamefont {Campbell}}]{PhysRevLett.106.130401}%
	\BibitemOpen
	\bibfield  {author} {\bibinfo {author} {\bibfnamefont {A.}~\bibnamefont {Ramanathan}}, \bibinfo {author} {\bibfnamefont {K.~C.}\ \bibnamefont {Wright}}, \bibinfo {author} {\bibfnamefont {S.~R.}\ \bibnamefont {Muniz}}, \bibinfo {author} {\bibfnamefont {M.}~\bibnamefont {Zelan}}, \bibinfo {author} {\bibfnamefont {W.~T.}\ \bibnamefont {Hill}}, \bibinfo {author} {\bibfnamefont {C.~J.}\ \bibnamefont {Lobb}}, \bibinfo {author} {\bibfnamefont {K.}~\bibnamefont {Helmerson}}, \bibinfo {author} {\bibfnamefont {W.~D.}\ \bibnamefont {Phillips}},\ and\ \bibinfo {author} {\bibfnamefont {G.~K.}\ \bibnamefont {Campbell}},\ }\bibfield  {title} {\bibinfo {title} {Superflow in a toroidal bose-einstein condensate: An atom circuit with a tunable weak link},\ }\href {https://doi.org/10.1103/PhysRevLett.106.130401} {\bibfield  {journal} {\bibinfo  {journal} {Phys. Rev. Lett.}\ }\textbf {\bibinfo {volume} {106}},\ \bibinfo {pages} {130401} (\bibinfo {year} {2011})}\BibitemShut {NoStop}%
	\bibitem [{\citenamefont {Marti}\ \emph {et~al.}(2015)\citenamefont {Marti}, \citenamefont {Olf},\ and\ \citenamefont {Stamper-Kurn}}]{PhysRevA.91.013602}%
	\BibitemOpen
	\bibfield  {author} {\bibinfo {author} {\bibfnamefont {G.~E.}\ \bibnamefont {Marti}}, \bibinfo {author} {\bibfnamefont {R.}~\bibnamefont {Olf}},\ and\ \bibinfo {author} {\bibfnamefont {D.~M.}\ \bibnamefont {Stamper-Kurn}},\ }\bibfield  {title} {\bibinfo {title} {Collective excitation interferometry with a toroidal bose-einstein condensate},\ }\href {https://doi.org/10.1103/PhysRevA.91.013602} {\bibfield  {journal} {\bibinfo  {journal} {Phys. Rev. A}\ }\textbf {\bibinfo {volume} {91}},\ \bibinfo {pages} {013602} (\bibinfo {year} {2015})}\BibitemShut {NoStop}%
	\bibitem [{\citenamefont {Ryu}\ \emph {et~al.}(2013)\citenamefont {Ryu}, \citenamefont {Blackburn}, \citenamefont {Blinova},\ and\ \citenamefont {Boshier}}]{PhysRevLett.111.205301}%
	\BibitemOpen
	\bibfield  {author} {\bibinfo {author} {\bibfnamefont {C.}~\bibnamefont {Ryu}}, \bibinfo {author} {\bibfnamefont {P.~W.}\ \bibnamefont {Blackburn}}, \bibinfo {author} {\bibfnamefont {A.~A.}\ \bibnamefont {Blinova}},\ and\ \bibinfo {author} {\bibfnamefont {M.~G.}\ \bibnamefont {Boshier}},\ }\bibfield  {title} {\bibinfo {title} {Experimental realization of josephson junctions for an atom squid},\ }\href {https://doi.org/10.1103/PhysRevLett.111.205301} {\bibfield  {journal} {\bibinfo  {journal} {Phys. Rev. Lett.}\ }\textbf {\bibinfo {volume} {111}},\ \bibinfo {pages} {205301} (\bibinfo {year} {2013})}\BibitemShut {NoStop}%
	\bibitem [{\citenamefont {Pandey}\ \emph {et~al.}(2021)\citenamefont {Pandey}, \citenamefont {Mas}, \citenamefont {Vasilakis},\ and\ \citenamefont {von Klitzing}}]{PhysRevLett.126.170402}%
	\BibitemOpen
	\bibfield  {author} {\bibinfo {author} {\bibfnamefont {S.}~\bibnamefont {Pandey}}, \bibinfo {author} {\bibfnamefont {H.}~\bibnamefont {Mas}}, \bibinfo {author} {\bibfnamefont {G.}~\bibnamefont {Vasilakis}},\ and\ \bibinfo {author} {\bibfnamefont {W.}~\bibnamefont {von Klitzing}},\ }\bibfield  {title} {\bibinfo {title} {Atomtronic matter-wave lensing},\ }\href {https://doi.org/10.1103/PhysRevLett.126.170402} {\bibfield  {journal} {\bibinfo  {journal} {Phys. Rev. Lett.}\ }\textbf {\bibinfo {volume} {126}},\ \bibinfo {pages} {170402} (\bibinfo {year} {2021})}\BibitemShut {NoStop}%
	\bibitem [{\citenamefont {Amico}\ \emph {et~al.}(2021)\citenamefont {Amico}, \citenamefont {Boshier}, \citenamefont {Birkl}, \citenamefont {Minguzzi}, \citenamefont {Miniatura}, \citenamefont {Kwek}, \citenamefont {Aghamalyan}, \citenamefont {Ahufinger}, \citenamefont {Anderson}, \citenamefont {Andrei}, \citenamefont {Arnold}, \citenamefont {Baker}, \citenamefont {Bell}, \citenamefont {Bland}, \citenamefont {Brantut} \emph {et~al.}}]{Amico2021}%
	\BibitemOpen
	\bibfield  {author} {\bibinfo {author} {\bibfnamefont {L.}~\bibnamefont {Amico}}, \bibinfo {author} {\bibfnamefont {M.}~\bibnamefont {Boshier}}, \bibinfo {author} {\bibfnamefont {G.}~\bibnamefont {Birkl}}, \bibinfo {author} {\bibfnamefont {A.}~\bibnamefont {Minguzzi}}, \bibinfo {author} {\bibfnamefont {C.}~\bibnamefont {Miniatura}}, \bibinfo {author} {\bibfnamefont {L.-C.}\ \bibnamefont {Kwek}}, \bibinfo {author} {\bibfnamefont {D.}~\bibnamefont {Aghamalyan}}, \bibinfo {author} {\bibfnamefont {V.}~\bibnamefont {Ahufinger}}, \bibinfo {author} {\bibfnamefont {D.}~\bibnamefont {Anderson}}, \bibinfo {author} {\bibfnamefont {N.}~\bibnamefont {Andrei}}, \bibinfo {author} {\bibfnamefont {A.~S.}\ \bibnamefont {Arnold}}, \bibinfo {author} {\bibfnamefont {M.}~\bibnamefont {Baker}}, \bibinfo {author} {\bibfnamefont {T.~A.}\ \bibnamefont {Bell}}, \bibinfo {author} {\bibfnamefont {T.}~\bibnamefont {Bland}}, \bibinfo {author} {\bibfnamefont {J.~P.}\ \bibnamefont {Brantut}}, \emph {et~al.},\ }\bibfield  {title} {\bibinfo
		{title} {Roadmap on atomtronics: State of the art and perspective},\ }\href {https://doi.org/10.1116/5.0026178} {\bibfield  {journal} {\bibinfo  {journal} {AVS Quantum Science}\ }\textbf {\bibinfo {volume} {3}},\ \bibinfo {pages} {039201} (\bibinfo {year} {2021})}\BibitemShut {NoStop}%
	\bibitem [{\citenamefont {Kanamoto}\ \emph {et~al.}(2008)\citenamefont {Kanamoto}, \citenamefont {Carr},\ and\ \citenamefont {Ueda}}]{PhysRevLett.100.060401}%
	\BibitemOpen
	\bibfield  {author} {\bibinfo {author} {\bibfnamefont {R.}~\bibnamefont {Kanamoto}}, \bibinfo {author} {\bibfnamefont {L.~D.}\ \bibnamefont {Carr}},\ and\ \bibinfo {author} {\bibfnamefont {M.}~\bibnamefont {Ueda}},\ }\bibfield  {title} {\bibinfo {title} {Topological winding and unwinding in metastable bose-einstein condensates},\ }\href {https://doi.org/10.1103/PhysRevLett.100.060401} {\bibfield  {journal} {\bibinfo  {journal} {Phys. Rev. Lett.}\ }\textbf {\bibinfo {volume} {100}},\ \bibinfo {pages} {060401} (\bibinfo {year} {2008})}\BibitemShut {NoStop}%
	\bibitem [{\citenamefont {Corman}\ \emph {et~al.}(2014)\citenamefont {Corman}, \citenamefont {Chomaz}, \citenamefont {Bienaim\'e}, \citenamefont {Desbuquois}, \citenamefont {Weitenberg}, \citenamefont {Nascimb\`ene}, \citenamefont {Dalibard},\ and\ \citenamefont {Beugnon}}]{PhysRevLett.113.135302}%
	\BibitemOpen
	\bibfield  {author} {\bibinfo {author} {\bibfnamefont {L.}~\bibnamefont {Corman}}, \bibinfo {author} {\bibfnamefont {L.}~\bibnamefont {Chomaz}}, \bibinfo {author} {\bibfnamefont {T.}~\bibnamefont {Bienaim\'e}}, \bibinfo {author} {\bibfnamefont {R.}~\bibnamefont {Desbuquois}}, \bibinfo {author} {\bibfnamefont {C.}~\bibnamefont {Weitenberg}}, \bibinfo {author} {\bibfnamefont {S.}~\bibnamefont {Nascimb\`ene}}, \bibinfo {author} {\bibfnamefont {J.}~\bibnamefont {Dalibard}},\ and\ \bibinfo {author} {\bibfnamefont {J.}~\bibnamefont {Beugnon}},\ }\bibfield  {title} {\bibinfo {title} {Quench-induced supercurrents in an annular bose gas},\ }\href {https://doi.org/10.1103/PhysRevLett.113.135302} {\bibfield  {journal} {\bibinfo  {journal} {Phys. Rev. Lett.}\ }\textbf {\bibinfo {volume} {113}},\ \bibinfo {pages} {135302} (\bibinfo {year} {2014})}\BibitemShut {NoStop}%
	\bibitem [{\citenamefont {Eckel}\ \emph {et~al.}(2014)\citenamefont {Eckel}, \citenamefont {Lee}, \citenamefont {Jendrzejewski}, \citenamefont {Murray}, \citenamefont {Clark}, \citenamefont {Lobb}, \citenamefont {Phillips}, \citenamefont {Edwards},\ and\ \citenamefont {Campbell}}]{Eckel2014}%
	\BibitemOpen
	\bibfield  {author} {\bibinfo {author} {\bibfnamefont {S.}~\bibnamefont {Eckel}}, \bibinfo {author} {\bibfnamefont {J.~G.}\ \bibnamefont {Lee}}, \bibinfo {author} {\bibfnamefont {F.}~\bibnamefont {Jendrzejewski}}, \bibinfo {author} {\bibfnamefont {N.}~\bibnamefont {Murray}}, \bibinfo {author} {\bibfnamefont {C.~W.}\ \bibnamefont {Clark}}, \bibinfo {author} {\bibfnamefont {C.~J.}\ \bibnamefont {Lobb}}, \bibinfo {author} {\bibfnamefont {W.~D.}\ \bibnamefont {Phillips}}, \bibinfo {author} {\bibfnamefont {M.}~\bibnamefont {Edwards}},\ and\ \bibinfo {author} {\bibfnamefont {G.~K.}\ \bibnamefont {Campbell}},\ }\bibfield  {title} {\bibinfo {title} {Hysteresis in a quantized superfluid `atomtronic' circuit},\ }\href {https://doi.org/10.1038/nature12958} {\bibfield  {journal} {\bibinfo  {journal} {Nature}\ }\textbf {\bibinfo {volume} {506}},\ \bibinfo {pages} {200} (\bibinfo {year} {2014})}\BibitemShut {NoStop}%
	\bibitem [{\citenamefont {Wang}\ \emph {et~al.}(2015)\citenamefont {Wang}, \citenamefont {Kumar}, \citenamefont {Jendrzejewski}, \citenamefont {Wilson}, \citenamefont {Edwards}, \citenamefont {Eckel}, \citenamefont {Campbell},\ and\ \citenamefont {Clark}}]{Wang_2015}%
	\BibitemOpen
	\bibfield  {author} {\bibinfo {author} {\bibfnamefont {Y.-H.}\ \bibnamefont {Wang}}, \bibinfo {author} {\bibfnamefont {A.}~\bibnamefont {Kumar}}, \bibinfo {author} {\bibfnamefont {F.}~\bibnamefont {Jendrzejewski}}, \bibinfo {author} {\bibfnamefont {R.~M.}\ \bibnamefont {Wilson}}, \bibinfo {author} {\bibfnamefont {M.}~\bibnamefont {Edwards}}, \bibinfo {author} {\bibfnamefont {S.}~\bibnamefont {Eckel}}, \bibinfo {author} {\bibfnamefont {G.~K.}\ \bibnamefont {Campbell}},\ and\ \bibinfo {author} {\bibfnamefont {C.~W.}\ \bibnamefont {Clark}},\ }\bibfield  {title} {\bibinfo {title} {Resonant wavepackets and shock waves in an atomtronic squid},\ }\href {https://doi.org/10.1088/1367-2630/17/12/125012} {\bibfield  {journal} {\bibinfo  {journal} {New Journal of Physics}\ }\textbf {\bibinfo {volume} {17}},\ \bibinfo {pages} {125012} (\bibinfo {year} {2015})}\BibitemShut {NoStop}%
	\bibitem [{\citenamefont {Wright}\ \emph {et~al.}(2013)\citenamefont {Wright}, \citenamefont {Blakestad}, \citenamefont {Lobb}, \citenamefont {Phillips},\ and\ \citenamefont {Campbell}}]{PhysRevLett.110.025302}%
	\BibitemOpen
	\bibfield  {author} {\bibinfo {author} {\bibfnamefont {K.~C.}\ \bibnamefont {Wright}}, \bibinfo {author} {\bibfnamefont {R.~B.}\ \bibnamefont {Blakestad}}, \bibinfo {author} {\bibfnamefont {C.~J.}\ \bibnamefont {Lobb}}, \bibinfo {author} {\bibfnamefont {W.~D.}\ \bibnamefont {Phillips}},\ and\ \bibinfo {author} {\bibfnamefont {G.~K.}\ \bibnamefont {Campbell}},\ }\bibfield  {title} {\bibinfo {title} {Driving phase slips in a superfluid atom circuit with a rotating weak link},\ }\href {https://doi.org/10.1103/PhysRevLett.110.025302} {\bibfield  {journal} {\bibinfo  {journal} {Phys. Rev. Lett.}\ }\textbf {\bibinfo {volume} {110}},\ \bibinfo {pages} {025302} (\bibinfo {year} {2013})}\BibitemShut {NoStop}%
	\bibitem [{\citenamefont {Moulder}\ \emph {et~al.}(2012)\citenamefont {Moulder}, \citenamefont {Beattie}, \citenamefont {Smith}, \citenamefont {Tammuz},\ and\ \citenamefont {Hadzibabic}}]{PhysRevA.86.013629}%
	\BibitemOpen
	\bibfield  {author} {\bibinfo {author} {\bibfnamefont {S.}~\bibnamefont {Moulder}}, \bibinfo {author} {\bibfnamefont {S.}~\bibnamefont {Beattie}}, \bibinfo {author} {\bibfnamefont {R.~P.}\ \bibnamefont {Smith}}, \bibinfo {author} {\bibfnamefont {N.}~\bibnamefont {Tammuz}},\ and\ \bibinfo {author} {\bibfnamefont {Z.}~\bibnamefont {Hadzibabic}},\ }\bibfield  {title} {\bibinfo {title} {Quantized supercurrent decay in an annular bose-einstein condensate},\ }\href {https://doi.org/10.1103/PhysRevA.86.013629} {\bibfield  {journal} {\bibinfo  {journal} {Phys. Rev. A}\ }\textbf {\bibinfo {volume} {86}},\ \bibinfo {pages} {013629} (\bibinfo {year} {2012})}\BibitemShut {NoStop}%
	\bibitem [{\citenamefont {\"Ohberg}\ and\ \citenamefont {Wright}(2019)}]{PhysRevLett.123.250402}%
	\BibitemOpen
	\bibfield  {author} {\bibinfo {author} {\bibfnamefont {P.}~\bibnamefont {\"Ohberg}}\ and\ \bibinfo {author} {\bibfnamefont {E.~M.}\ \bibnamefont {Wright}},\ }\bibfield  {title} {\bibinfo {title} {Quantum time crystals and interacting gauge theories in atomic bose-einstein condensates},\ }\href {https://doi.org/10.1103/PhysRevLett.123.250402} {\bibfield  {journal} {\bibinfo  {journal} {Phys. Rev. Lett.}\ }\textbf {\bibinfo {volume} {123}},\ \bibinfo {pages} {250402} (\bibinfo {year} {2019})}\BibitemShut {NoStop}%
	\bibitem [{\citenamefont {Cooper}\ \emph {et~al.}(2010)\citenamefont {Cooper}, \citenamefont {Hallwood},\ and\ \citenamefont {Dunningham}}]{PhysRevA.81.043624}%
	\BibitemOpen
	\bibfield  {author} {\bibinfo {author} {\bibfnamefont {J.~J.}\ \bibnamefont {Cooper}}, \bibinfo {author} {\bibfnamefont {D.~W.}\ \bibnamefont {Hallwood}},\ and\ \bibinfo {author} {\bibfnamefont {J.~A.}\ \bibnamefont {Dunningham}},\ }\bibfield  {title} {\bibinfo {title} {Entanglement-enhanced atomic gyroscope},\ }\href {https://doi.org/10.1103/PhysRevA.81.043624} {\bibfield  {journal} {\bibinfo  {journal} {Phys. Rev. A}\ }\textbf {\bibinfo {volume} {81}},\ \bibinfo {pages} {043624} (\bibinfo {year} {2010})}\BibitemShut {NoStop}%
	\bibitem [{\citenamefont {Eckel}\ \emph {et~al.}(2018)\citenamefont {Eckel}, \citenamefont {Kumar}, \citenamefont {Jacobson}, \citenamefont {Spielman},\ and\ \citenamefont {Campbell}}]{PhysRevX.8.021021}%
	\BibitemOpen
	\bibfield  {author} {\bibinfo {author} {\bibfnamefont {S.}~\bibnamefont {Eckel}}, \bibinfo {author} {\bibfnamefont {A.}~\bibnamefont {Kumar}}, \bibinfo {author} {\bibfnamefont {T.}~\bibnamefont {Jacobson}}, \bibinfo {author} {\bibfnamefont {I.~B.}\ \bibnamefont {Spielman}},\ and\ \bibinfo {author} {\bibfnamefont {G.~K.}\ \bibnamefont {Campbell}},\ }\bibfield  {title} {\bibinfo {title} {A rapidly expanding bose-einstein condensate: An expanding universe in the lab},\ }\href {https://doi.org/10.1103/PhysRevX.8.021021} {\bibfield  {journal} {\bibinfo  {journal} {Phys. Rev. X}\ }\textbf {\bibinfo {volume} {8}},\ \bibinfo {pages} {021021} (\bibinfo {year} {2018})}\BibitemShut {NoStop}%
	\bibitem [{\citenamefont {Kumar}\ \emph {et~al.}(2016)\citenamefont {Kumar}, \citenamefont {Anderson}, \citenamefont {Phillips}, \citenamefont {Eckel}, \citenamefont {Campbell},\ and\ \citenamefont {Stringari}}]{Kumar_2016}%
	\BibitemOpen
	\bibfield  {author} {\bibinfo {author} {\bibfnamefont {A.}~\bibnamefont {Kumar}}, \bibinfo {author} {\bibfnamefont {N.}~\bibnamefont {Anderson}}, \bibinfo {author} {\bibfnamefont {W.~D.}\ \bibnamefont {Phillips}}, \bibinfo {author} {\bibfnamefont {S.}~\bibnamefont {Eckel}}, \bibinfo {author} {\bibfnamefont {G.~K.}\ \bibnamefont {Campbell}},\ and\ \bibinfo {author} {\bibfnamefont {S.}~\bibnamefont {Stringari}},\ }\bibfield  {title} {\bibinfo {title} {Minimally destructive, doppler measurement of a quantized flow in a ring-shaped bose–einstein condensate},\ }\href {https://doi.org/10.1088/1367-2630/18/2/025001} {\bibfield  {journal} {\bibinfo  {journal} {New Journal of Physics}\ }\textbf {\bibinfo {volume} {18}},\ \bibinfo {pages} {025001} (\bibinfo {year} {2016})}\BibitemShut {NoStop}%
	\bibitem [{\citenamefont {Freilich}\ \emph {et~al.}(2010)\citenamefont {Freilich}, \citenamefont {Bianchi}, \citenamefont {Kaufman}, \citenamefont {Langin},\ and\ \citenamefont {Hall}}]{doi:10.1126/science.1191224}%
	\BibitemOpen
	\bibfield  {author} {\bibinfo {author} {\bibfnamefont {D.~V.}\ \bibnamefont {Freilich}}, \bibinfo {author} {\bibfnamefont {D.~M.}\ \bibnamefont {Bianchi}}, \bibinfo {author} {\bibfnamefont {A.~M.}\ \bibnamefont {Kaufman}}, \bibinfo {author} {\bibfnamefont {T.~K.}\ \bibnamefont {Langin}},\ and\ \bibinfo {author} {\bibfnamefont {D.~S.}\ \bibnamefont {Hall}},\ }\bibfield  {title} {\bibinfo {title} {Real-time dynamics of single vortex lines and vortex dipoles in a bose-einstein condensate},\ }\href {https://doi.org/10.1126/science.1191224} {\bibfield  {journal} {\bibinfo  {journal} {Science}\ }\textbf {\bibinfo {volume} {329}},\ \bibinfo {pages} {1182} (\bibinfo {year} {2010})},\ \Eprint {https://arxiv.org/abs/https://www.science.org/doi/pdf/10.1126/science.1191224} {https://www.science.org/doi/pdf/10.1126/science.1191224} \BibitemShut {NoStop}%
	\bibitem [{\citenamefont {Kumar}\ \emph {et~al.}(2021)\citenamefont {Kumar}, \citenamefont {Biswas}, \citenamefont {Feliz}, \citenamefont {Kanamoto}, \citenamefont {Chang}, \citenamefont {Jha},\ and\ \citenamefont {Bhattacharya}}]{PhysRevLett.127.113601}%
	\BibitemOpen
	\bibfield  {author} {\bibinfo {author} {\bibfnamefont {P.}~\bibnamefont {Kumar}}, \bibinfo {author} {\bibfnamefont {T.}~\bibnamefont {Biswas}}, \bibinfo {author} {\bibfnamefont {K.}~\bibnamefont {Feliz}}, \bibinfo {author} {\bibfnamefont {R.}~\bibnamefont {Kanamoto}}, \bibinfo {author} {\bibfnamefont {M.-S.}\ \bibnamefont {Chang}}, \bibinfo {author} {\bibfnamefont {A.~K.}\ \bibnamefont {Jha}},\ and\ \bibinfo {author} {\bibfnamefont {M.}~\bibnamefont {Bhattacharya}},\ }\bibfield  {title} {\bibinfo {title} {Cavity optomechanical sensing and manipulation of an atomic persistent current},\ }\href {https://doi.org/10.1103/PhysRevLett.127.113601} {\bibfield  {journal} {\bibinfo  {journal} {Phys. Rev. Lett.}\ }\textbf {\bibinfo {volume} {127}},\ \bibinfo {pages} {113601} (\bibinfo {year} {2021})}\BibitemShut {NoStop}%
	\bibitem [{\citenamefont {Pradhan}\ \emph {et~al.}(2024{\natexlab{a}})\citenamefont {Pradhan}, \citenamefont {Kumar}, \citenamefont {Kanamoto}, \citenamefont {Dey}, \citenamefont {Bhattacharya},\ and\ \citenamefont {Mishra}}]{Pradhan_PRR2024}%
	\BibitemOpen
	\bibfield  {author} {\bibinfo {author} {\bibfnamefont {N.}~\bibnamefont {Pradhan}}, \bibinfo {author} {\bibfnamefont {P.}~\bibnamefont {Kumar}}, \bibinfo {author} {\bibfnamefont {R.}~\bibnamefont {Kanamoto}}, \bibinfo {author} {\bibfnamefont {T.~N.}\ \bibnamefont {Dey}}, \bibinfo {author} {\bibfnamefont {M.}~\bibnamefont {Bhattacharya}},\ and\ \bibinfo {author} {\bibfnamefont {P.~K.}\ \bibnamefont {Mishra}},\ }\bibfield  {title} {\bibinfo {title} {Cavity optomechanical detection of persistent currents and solitons in a bosonic ring condensate},\ }\href {https://doi.org/10.1103/PhysRevResearch.6.013104} {\bibfield  {journal} {\bibinfo  {journal} {Phys. Rev. Res.}\ }\textbf {\bibinfo {volume} {6}},\ \bibinfo {pages} {013104} (\bibinfo {year} {2024}{\natexlab{a}})}\BibitemShut {NoStop}%
	\bibitem [{\citenamefont {Pradhan}\ \emph {et~al.}(2024{\natexlab{b}})\citenamefont {Pradhan}, \citenamefont {Kumar}, \citenamefont {Kanamoto}, \citenamefont {Dey}, \citenamefont {Bhattacharya},\ and\ \citenamefont {Mishra}}]{Pradhan_PRA2024}%
	\BibitemOpen
	\bibfield  {author} {\bibinfo {author} {\bibfnamefont {N.}~\bibnamefont {Pradhan}}, \bibinfo {author} {\bibfnamefont {P.}~\bibnamefont {Kumar}}, \bibinfo {author} {\bibfnamefont {R.}~\bibnamefont {Kanamoto}}, \bibinfo {author} {\bibfnamefont {T.~N.}\ \bibnamefont {Dey}}, \bibinfo {author} {\bibfnamefont {M.}~\bibnamefont {Bhattacharya}},\ and\ \bibinfo {author} {\bibfnamefont {P.~K.}\ \bibnamefont {Mishra}},\ }\bibfield  {title} {\bibinfo {title} {Ring bose-einstein condensate in a cavity: Chirality detection and rotation sensing},\ }\href {https://doi.org/10.1103/PhysRevA.109.023524} {\bibfield  {journal} {\bibinfo  {journal} {Phys. Rev. A}\ }\textbf {\bibinfo {volume} {109}},\ \bibinfo {pages} {023524} (\bibinfo {year} {2024}{\natexlab{b}})}\BibitemShut {NoStop}%
	\bibitem [{\citenamefont {Aspelmeyer}\ \emph {et~al.}(2014)\citenamefont {Aspelmeyer}, \citenamefont {Kippenberg},\ and\ \citenamefont {Marquardt}}]{RevModPhys.86.1391}%
	\BibitemOpen
	\bibfield  {author} {\bibinfo {author} {\bibfnamefont {M.}~\bibnamefont {Aspelmeyer}}, \bibinfo {author} {\bibfnamefont {T.~J.}\ \bibnamefont {Kippenberg}},\ and\ \bibinfo {author} {\bibfnamefont {F.}~\bibnamefont {Marquardt}},\ }\bibfield  {title} {\bibinfo {title} {Cavity optomechanics},\ }\href {https://doi.org/10.1103/RevModPhys.86.1391} {\bibfield  {journal} {\bibinfo  {journal} {Rev. Mod. Phys.}\ }\textbf {\bibinfo {volume} {86}},\ \bibinfo {pages} {1391} (\bibinfo {year} {2014})}\BibitemShut {NoStop}%
	\bibitem [{\citenamefont {Genes}\ \emph {et~al.}(2008{\natexlab{a}})\citenamefont {Genes}, \citenamefont {Vitali}, \citenamefont {Tombesi}, \citenamefont {Gigan},\ and\ \citenamefont {Aspelmeyer}}]{PhysRevA.77.033804}%
	\BibitemOpen
	\bibfield  {author} {\bibinfo {author} {\bibfnamefont {C.}~\bibnamefont {Genes}}, \bibinfo {author} {\bibfnamefont {D.}~\bibnamefont {Vitali}}, \bibinfo {author} {\bibfnamefont {P.}~\bibnamefont {Tombesi}}, \bibinfo {author} {\bibfnamefont {S.}~\bibnamefont {Gigan}},\ and\ \bibinfo {author} {\bibfnamefont {M.}~\bibnamefont {Aspelmeyer}},\ }\bibfield  {title} {\bibinfo {title} {Ground-state cooling of a micromechanical oscillator: Comparing cold damping and cavity-assisted cooling schemes},\ }\href {https://doi.org/10.1103/PhysRevA.77.033804} {\bibfield  {journal} {\bibinfo  {journal} {Phys. Rev. A}\ }\textbf {\bibinfo {volume} {77}},\ \bibinfo {pages} {033804} (\bibinfo {year} {2008}{\natexlab{a}})}\BibitemShut {NoStop}%
	\bibitem [{\citenamefont {Marquardt}\ \emph {et~al.}(2007)\citenamefont {Marquardt}, \citenamefont {Chen}, \citenamefont {Clerk},\ and\ \citenamefont {Girvin}}]{PhysRevLett.99.093902}%
	\BibitemOpen
	\bibfield  {author} {\bibinfo {author} {\bibfnamefont {F.}~\bibnamefont {Marquardt}}, \bibinfo {author} {\bibfnamefont {J.~P.}\ \bibnamefont {Chen}}, \bibinfo {author} {\bibfnamefont {A.~A.}\ \bibnamefont {Clerk}},\ and\ \bibinfo {author} {\bibfnamefont {S.~M.}\ \bibnamefont {Girvin}},\ }\bibfield  {title} {\bibinfo {title} {Quantum theory of cavity-assisted sideband cooling of mechanical motion},\ }\href {https://doi.org/10.1103/PhysRevLett.99.093902} {\bibfield  {journal} {\bibinfo  {journal} {Phys. Rev. Lett.}\ }\textbf {\bibinfo {volume} {99}},\ \bibinfo {pages} {093902} (\bibinfo {year} {2007})}\BibitemShut {NoStop}%
	\bibitem [{\citenamefont {Wilson-Rae}\ \emph {et~al.}(2007)\citenamefont {Wilson-Rae}, \citenamefont {Nooshi}, \citenamefont {Zwerger},\ and\ \citenamefont {Kippenberg}}]{PhysRevLett.99.093901}%
	\BibitemOpen
	\bibfield  {author} {\bibinfo {author} {\bibfnamefont {I.}~\bibnamefont {Wilson-Rae}}, \bibinfo {author} {\bibfnamefont {N.}~\bibnamefont {Nooshi}}, \bibinfo {author} {\bibfnamefont {W.}~\bibnamefont {Zwerger}},\ and\ \bibinfo {author} {\bibfnamefont {T.~J.}\ \bibnamefont {Kippenberg}},\ }\bibfield  {title} {\bibinfo {title} {Theory of ground state cooling of a mechanical oscillator using dynamical backaction},\ }\href {https://doi.org/10.1103/PhysRevLett.99.093901} {\bibfield  {journal} {\bibinfo  {journal} {Phys. Rev. Lett.}\ }\textbf {\bibinfo {volume} {99}},\ \bibinfo {pages} {093901} (\bibinfo {year} {2007})}\BibitemShut {NoStop}%
	\bibitem [{\citenamefont {Mancini}\ \emph {et~al.}(2002)\citenamefont {Mancini}, \citenamefont {Giovannetti}, \citenamefont {Vitali},\ and\ \citenamefont {Tombesi}}]{PhysRevLett.88.120401}%
	\BibitemOpen
	\bibfield  {author} {\bibinfo {author} {\bibfnamefont {S.}~\bibnamefont {Mancini}}, \bibinfo {author} {\bibfnamefont {V.}~\bibnamefont {Giovannetti}}, \bibinfo {author} {\bibfnamefont {D.}~\bibnamefont {Vitali}},\ and\ \bibinfo {author} {\bibfnamefont {P.}~\bibnamefont {Tombesi}},\ }\bibfield  {title} {\bibinfo {title} {Entangling macroscopic oscillators exploiting radiation pressure},\ }\href {https://doi.org/10.1103/PhysRevLett.88.120401} {\bibfield  {journal} {\bibinfo  {journal} {Phys. Rev. Lett.}\ }\textbf {\bibinfo {volume} {88}},\ \bibinfo {pages} {120401} (\bibinfo {year} {2002})}\BibitemShut {NoStop}%
	\bibitem [{\citenamefont {Vitali}\ \emph {et~al.}(2007)\citenamefont {Vitali}, \citenamefont {Mancini},\ and\ \citenamefont {Tombesi}}]{Vitali_2007}%
	\BibitemOpen
	\bibfield  {author} {\bibinfo {author} {\bibfnamefont {D.}~\bibnamefont {Vitali}}, \bibinfo {author} {\bibfnamefont {S.}~\bibnamefont {Mancini}},\ and\ \bibinfo {author} {\bibfnamefont {P.}~\bibnamefont {Tombesi}},\ }\bibfield  {title} {\bibinfo {title} {Stationary entanglement between two movable mirrors in a classically driven fabry–perot cavity},\ }\href {https://doi.org/10.1088/1751-8113/40/28/S14} {\bibfield  {journal} {\bibinfo  {journal} {Journal of Physics A: Mathematical and Theoretical}\ }\textbf {\bibinfo {volume} {40}},\ \bibinfo {pages} {8055} (\bibinfo {year} {2007})}\BibitemShut {NoStop}%
	\bibitem [{\citenamefont {Paternostro}\ \emph {et~al.}(2007)\citenamefont {Paternostro}, \citenamefont {Vitali}, \citenamefont {Gigan}, \citenamefont {Kim}, \citenamefont {Brukner}, \citenamefont {Eisert},\ and\ \citenamefont {Aspelmeyer}}]{PhysRevLett.99.250401}%
	\BibitemOpen
	\bibfield  {author} {\bibinfo {author} {\bibfnamefont {M.}~\bibnamefont {Paternostro}}, \bibinfo {author} {\bibfnamefont {D.}~\bibnamefont {Vitali}}, \bibinfo {author} {\bibfnamefont {S.}~\bibnamefont {Gigan}}, \bibinfo {author} {\bibfnamefont {M.~S.}\ \bibnamefont {Kim}}, \bibinfo {author} {\bibfnamefont {C.}~\bibnamefont {Brukner}}, \bibinfo {author} {\bibfnamefont {J.}~\bibnamefont {Eisert}},\ and\ \bibinfo {author} {\bibfnamefont {M.}~\bibnamefont {Aspelmeyer}},\ }\bibfield  {title} {\bibinfo {title} {Creating and probing multipartite macroscopic entanglement with light},\ }\href {https://doi.org/10.1103/PhysRevLett.99.250401} {\bibfield  {journal} {\bibinfo  {journal} {Phys. Rev. Lett.}\ }\textbf {\bibinfo {volume} {99}},\ \bibinfo {pages} {250401} (\bibinfo {year} {2007})}\BibitemShut {NoStop}%
	\bibitem [{\citenamefont {Genes}\ \emph {et~al.}(2008{\natexlab{b}})\citenamefont {Genes}, \citenamefont {Vitali},\ and\ \citenamefont {Tombesi}}]{Genes_2008}%
	\BibitemOpen
	\bibfield  {author} {\bibinfo {author} {\bibfnamefont {C.}~\bibnamefont {Genes}}, \bibinfo {author} {\bibfnamefont {D.}~\bibnamefont {Vitali}},\ and\ \bibinfo {author} {\bibfnamefont {P.}~\bibnamefont {Tombesi}},\ }\bibfield  {title} {\bibinfo {title} {Simultaneous cooling and entanglement of mechanical modes of a micromirror in an optical cavity},\ }\href {https://doi.org/10.1088/1367-2630/10/9/095009} {\bibfield  {journal} {\bibinfo  {journal} {New Journal of Physics}\ }\textbf {\bibinfo {volume} {10}},\ \bibinfo {pages} {095009} (\bibinfo {year} {2008}{\natexlab{b}})}\BibitemShut {NoStop}%
	\bibitem [{\citenamefont {Fabre}\ \emph {et~al.}(1994{\natexlab{a}})\citenamefont {Fabre}, \citenamefont {Pinard}, \citenamefont {Bourzeix}, \citenamefont {Heidmann}, \citenamefont {Giacobino},\ and\ \citenamefont {Reynaud}}]{PhysRevA.49.1337}%
	\BibitemOpen
	\bibfield  {author} {\bibinfo {author} {\bibfnamefont {C.}~\bibnamefont {Fabre}}, \bibinfo {author} {\bibfnamefont {M.}~\bibnamefont {Pinard}}, \bibinfo {author} {\bibfnamefont {S.}~\bibnamefont {Bourzeix}}, \bibinfo {author} {\bibfnamefont {A.}~\bibnamefont {Heidmann}}, \bibinfo {author} {\bibfnamefont {E.}~\bibnamefont {Giacobino}},\ and\ \bibinfo {author} {\bibfnamefont {S.}~\bibnamefont {Reynaud}},\ }\bibfield  {title} {\bibinfo {title} {Quantum-noise reduction using a cavity with a movable mirror},\ }\href {https://doi.org/10.1103/PhysRevA.49.1337} {\bibfield  {journal} {\bibinfo  {journal} {Phys. Rev. A}\ }\textbf {\bibinfo {volume} {49}},\ \bibinfo {pages} {1337} (\bibinfo {year} {1994}{\natexlab{a}})}\BibitemShut {NoStop}%
	\bibitem [{\citenamefont {Mancini}\ and\ \citenamefont {Tombesi}(1994{\natexlab{a}})}]{PhysRevA.49.4055}%
	\BibitemOpen
	\bibfield  {author} {\bibinfo {author} {\bibfnamefont {S.}~\bibnamefont {Mancini}}\ and\ \bibinfo {author} {\bibfnamefont {P.}~\bibnamefont {Tombesi}},\ }\bibfield  {title} {\bibinfo {title} {Quantum noise reduction by radiation pressure},\ }\href {https://doi.org/10.1103/PhysRevA.49.4055} {\bibfield  {journal} {\bibinfo  {journal} {Phys. Rev. A}\ }\textbf {\bibinfo {volume} {49}},\ \bibinfo {pages} {4055} (\bibinfo {year} {1994}{\natexlab{a}})}\BibitemShut {NoStop}%
	\bibitem [{\citenamefont {Qu}\ and\ \citenamefont {Agarwal}(2014)}]{Qu_2014}%
	\BibitemOpen
	\bibfield  {author} {\bibinfo {author} {\bibfnamefont {K.}~\bibnamefont {Qu}}\ and\ \bibinfo {author} {\bibfnamefont {G.~S.}\ \bibnamefont {Agarwal}},\ }\bibfield  {title} {\bibinfo {title} {Strong squeezing via phonon mediated spontaneous generation of photon pairs},\ }\href {https://doi.org/10.1088/1367-2630/16/11/113004} {\bibfield  {journal} {\bibinfo  {journal} {New Journal of Physics}\ }\textbf {\bibinfo {volume} {16}},\ \bibinfo {pages} {113004} (\bibinfo {year} {2014})}\BibitemShut {NoStop}%
	\bibitem [{\citenamefont {Aasi}\ \emph {et~al.}(2013)\citenamefont {Aasi}, \citenamefont {Abadie}, \citenamefont {Abbott}, \citenamefont {Abbott}, \citenamefont {Abbott}, \citenamefont {Abernathy}, \citenamefont {Adams}, \citenamefont {Adams}, \citenamefont {Addesso}, \citenamefont {Adhikari}, \citenamefont {Affeldt}, \citenamefont {Aguiar}, \citenamefont {Ajith}, \citenamefont {Allen}, \citenamefont {Amador~Ceron} \emph {et~al.}}]{Aasi_NatPhoton2013}%
	\BibitemOpen
	\bibfield  {author} {\bibinfo {author} {\bibfnamefont {J.}~\bibnamefont {Aasi}}, \bibinfo {author} {\bibfnamefont {J.}~\bibnamefont {Abadie}}, \bibinfo {author} {\bibfnamefont {B.~P.}\ \bibnamefont {Abbott}}, \bibinfo {author} {\bibfnamefont {R.}~\bibnamefont {Abbott}}, \bibinfo {author} {\bibfnamefont {T.~D.}\ \bibnamefont {Abbott}}, \bibinfo {author} {\bibfnamefont {M.~R.}\ \bibnamefont {Abernathy}}, \bibinfo {author} {\bibfnamefont {C.}~\bibnamefont {Adams}}, \bibinfo {author} {\bibfnamefont {T.}~\bibnamefont {Adams}}, \bibinfo {author} {\bibfnamefont {P.}~\bibnamefont {Addesso}}, \bibinfo {author} {\bibfnamefont {R.~X.}\ \bibnamefont {Adhikari}}, \bibinfo {author} {\bibfnamefont {C.}~\bibnamefont {Affeldt}}, \bibinfo {author} {\bibfnamefont {O.~D.}\ \bibnamefont {Aguiar}}, \bibinfo {author} {\bibfnamefont {P.}~\bibnamefont {Ajith}}, \bibinfo {author} {\bibfnamefont {B.}~\bibnamefont {Allen}}, \bibinfo {author} {\bibfnamefont {E.}~\bibnamefont {Amador~Ceron}}, \emph {et~al.},\ }\bibfield  {title}
	{\bibinfo {title} {Enhanced sensitivity of the ligo gravitational wave detector by using squeezed states of light},\ }\href {https://doi.org/10.1038/nphoton.2013.177} {\bibfield  {journal} {\bibinfo  {journal} {Nature Photonics}\ }\textbf {\bibinfo {volume} {7}},\ \bibinfo {pages} {613} (\bibinfo {year} {2013})}\BibitemShut {NoStop}%
	\bibitem [{\citenamefont {Ganapathy}\ \emph {et~al.}(2023)\citenamefont {Ganapathy}, \citenamefont {Jia}, \citenamefont {Nakano}, \citenamefont {Xu}, \citenamefont {Aritomi}, \citenamefont {Cullen}, \citenamefont {Kijbunchoo}, \citenamefont {Dwyer}, \citenamefont {Mullavey}, \citenamefont {McCuller}, \citenamefont {Abbott}, \citenamefont {Abouelfettouh}, \citenamefont {Adhikari}, \citenamefont {Ananyeva}, \citenamefont {Appert} \emph {et~al.}}]{Ganapathy_PRX2023}%
	\BibitemOpen
	\bibfield  {author} {\bibinfo {author} {\bibfnamefont {D.}~\bibnamefont {Ganapathy}}, \bibinfo {author} {\bibfnamefont {W.}~\bibnamefont {Jia}}, \bibinfo {author} {\bibfnamefont {M.}~\bibnamefont {Nakano}}, \bibinfo {author} {\bibfnamefont {V.}~\bibnamefont {Xu}}, \bibinfo {author} {\bibfnamefont {N.}~\bibnamefont {Aritomi}}, \bibinfo {author} {\bibfnamefont {T.}~\bibnamefont {Cullen}}, \bibinfo {author} {\bibfnamefont {N.}~\bibnamefont {Kijbunchoo}}, \bibinfo {author} {\bibfnamefont {S.~E.}\ \bibnamefont {Dwyer}}, \bibinfo {author} {\bibfnamefont {A.}~\bibnamefont {Mullavey}}, \bibinfo {author} {\bibfnamefont {L.}~\bibnamefont {McCuller}}, \bibinfo {author} {\bibfnamefont {R.}~\bibnamefont {Abbott}}, \bibinfo {author} {\bibfnamefont {I.}~\bibnamefont {Abouelfettouh}}, \bibinfo {author} {\bibfnamefont {R.~X.}\ \bibnamefont {Adhikari}}, \bibinfo {author} {\bibfnamefont {A.}~\bibnamefont {Ananyeva}}, \bibinfo {author} {\bibfnamefont {S.}~\bibnamefont {Appert}}, \emph {et~al.} (\bibinfo {collaboration} {LIGO
			O4 Detector Collaboration}),\ }\bibfield  {title} {\bibinfo {title} {Broadband quantum enhancement of the ligo detectors with frequency-dependent squeezing},\ }\href {https://doi.org/10.1103/PhysRevX.13.041021} {\bibfield  {journal} {\bibinfo  {journal} {Phys. Rev. X}\ }\textbf {\bibinfo {volume} {13}},\ \bibinfo {pages} {041021} (\bibinfo {year} {2023})}\BibitemShut {NoStop}%
	\bibitem [{\citenamefont {Peuntinger}\ \emph {et~al.}(2014)\citenamefont {Peuntinger}, \citenamefont {Heim}, \citenamefont {M\"uller}, \citenamefont {Gabriel}, \citenamefont {Marquardt},\ and\ \citenamefont {Leuchs}}]{Peuntinger_PRL2014}%
	\BibitemOpen
	\bibfield  {author} {\bibinfo {author} {\bibfnamefont {C.}~\bibnamefont {Peuntinger}}, \bibinfo {author} {\bibfnamefont {B.}~\bibnamefont {Heim}}, \bibinfo {author} {\bibfnamefont {C.~R.}\ \bibnamefont {M\"uller}}, \bibinfo {author} {\bibfnamefont {C.}~\bibnamefont {Gabriel}}, \bibinfo {author} {\bibfnamefont {C.}~\bibnamefont {Marquardt}},\ and\ \bibinfo {author} {\bibfnamefont {G.}~\bibnamefont {Leuchs}},\ }\bibfield  {title} {\bibinfo {title} {Distribution of squeezed states through an atmospheric channel},\ }\href {https://doi.org/10.1103/PhysRevLett.113.060502} {\bibfield  {journal} {\bibinfo  {journal} {Phys. Rev. Lett.}\ }\textbf {\bibinfo {volume} {113}},\ \bibinfo {pages} {060502} (\bibinfo {year} {2014})}\BibitemShut {NoStop}%
	\bibitem [{\citenamefont {Lee}\ \emph {et~al.}(2020)\citenamefont {Lee}, \citenamefont {Lee},\ and\ \citenamefont {Seok}}]{Lee_SciRep2020}%
	\BibitemOpen
	\bibfield  {author} {\bibinfo {author} {\bibfnamefont {C.-W.}\ \bibnamefont {Lee}}, \bibinfo {author} {\bibfnamefont {J.~H.}\ \bibnamefont {Lee}},\ and\ \bibinfo {author} {\bibfnamefont {H.}~\bibnamefont {Seok}},\ }\bibfield  {title} {\bibinfo {title} {Squeezed-light-driven force detection with an optomechanical cavity in a mach{\^a}{\texteuro}``zehnder interferometer},\ }\href {https://doi.org/10.1038/s41598-020-74629-1} {\bibfield  {journal} {\bibinfo  {journal} {Scientific Reports}\ }\textbf {\bibinfo {volume} {10}},\ \bibinfo {pages} {17496} (\bibinfo {year} {2020})}\BibitemShut {NoStop}%
	\bibitem [{\citenamefont {Lawrie}\ \emph {et~al.}(2019)\citenamefont {Lawrie}, \citenamefont {Lett}, \citenamefont {Marino},\ and\ \citenamefont {Pooser}}]{Lawrie_ACSPhoton2019}%
	\BibitemOpen
	\bibfield  {author} {\bibinfo {author} {\bibfnamefont {B.~J.}\ \bibnamefont {Lawrie}}, \bibinfo {author} {\bibfnamefont {P.~D.}\ \bibnamefont {Lett}}, \bibinfo {author} {\bibfnamefont {A.~M.}\ \bibnamefont {Marino}},\ and\ \bibinfo {author} {\bibfnamefont {R.~C.}\ \bibnamefont {Pooser}},\ }\bibfield  {title} {\bibinfo {title} {Quantum sensing with squeezed light},\ }\href {https://doi.org/10.1021/acsphotonics.9b00250} {\bibfield  {journal} {\bibinfo  {journal} {ACS Photonics}\ }\textbf {\bibinfo {volume} {6}},\ \bibinfo {pages} {1307} (\bibinfo {year} {2019})}\BibitemShut {NoStop}%
	\bibitem [{\citenamefont {Shi}\ and\ \citenamefont {Bhattacharya}(2016)}]{Shi_JPB2016}%
	\BibitemOpen
	\bibfield  {author} {\bibinfo {author} {\bibfnamefont {H.}~\bibnamefont {Shi}}\ and\ \bibinfo {author} {\bibfnamefont {M.}~\bibnamefont {Bhattacharya}},\ }\bibfield  {title} {\bibinfo {title} {Optomechanics based on angular momentum exchange between light and matter},\ }\href {https://doi.org/10.1088/0953-4075/49/15/153001} {\bibfield  {journal} {\bibinfo  {journal} {Journal of Physics B: Atomic, Molecular and Optical Physics}\ }\textbf {\bibinfo {volume} {49}},\ \bibinfo {pages} {153001} (\bibinfo {year} {2016})}\BibitemShut {NoStop}%
	\bibitem [{\citenamefont {Bhattacharya}(2015)}]{Bhattacharya_JOSAB2015}%
	\BibitemOpen
	\bibfield  {author} {\bibinfo {author} {\bibfnamefont {M.}~\bibnamefont {Bhattacharya}},\ }\bibfield  {title} {\bibinfo {title} {Rotational cavity optomechanics},\ }\href {https://doi.org/10.1364/JOSAB.32.000B55} {\bibfield  {journal} {\bibinfo  {journal} {J. Opt. Soc. Am. B}\ }\textbf {\bibinfo {volume} {32}},\ \bibinfo {pages} {B55} (\bibinfo {year} {2015})}\BibitemShut {NoStop}%
	\bibitem [{\citenamefont {Bhattacharya}\ and\ \citenamefont {Meystre}(2007)}]{Bhattacharya_PRL2007}%
	\BibitemOpen
	\bibfield  {author} {\bibinfo {author} {\bibfnamefont {M.}~\bibnamefont {Bhattacharya}}\ and\ \bibinfo {author} {\bibfnamefont {P.}~\bibnamefont {Meystre}},\ }\bibfield  {title} {\bibinfo {title} {Using a laguerre-gaussian beam to trap and cool the rotational motion of a mirror},\ }\href {https://doi.org/10.1103/PhysRevLett.99.153603} {\bibfield  {journal} {\bibinfo  {journal} {Phys. Rev. Lett.}\ }\textbf {\bibinfo {volume} {99}},\ \bibinfo {pages} {153603} (\bibinfo {year} {2007})}\BibitemShut {NoStop}%
	\bibitem [{\citenamefont {Bhattacharya}\ \emph {et~al.}(2008{\natexlab{a}})\citenamefont {Bhattacharya}, \citenamefont {Giscard},\ and\ \citenamefont {Meystre}}]{Bhattacharya_PRA2008}%
	\BibitemOpen
	\bibfield  {author} {\bibinfo {author} {\bibfnamefont {M.}~\bibnamefont {Bhattacharya}}, \bibinfo {author} {\bibfnamefont {P.-L.}\ \bibnamefont {Giscard}},\ and\ \bibinfo {author} {\bibfnamefont {P.}~\bibnamefont {Meystre}},\ }\bibfield  {title} {\bibinfo {title} {Entanglement of a laguerre-gaussian cavity mode with a rotating mirror},\ }\href {https://doi.org/10.1103/PhysRevA.77.013827} {\bibfield  {journal} {\bibinfo  {journal} {Phys. Rev. A}\ }\textbf {\bibinfo {volume} {77}},\ \bibinfo {pages} {013827} (\bibinfo {year} {2008}{\natexlab{a}})}\BibitemShut {NoStop}%
	\bibitem [{\citenamefont {Bhattacharya}\ \emph {et~al.}(2008{\natexlab{b}})\citenamefont {Bhattacharya}, \citenamefont {Giscard},\ and\ \citenamefont {Meystre}}]{Bhattacharya_PRAII2008}%
	\BibitemOpen
	\bibfield  {author} {\bibinfo {author} {\bibfnamefont {M.}~\bibnamefont {Bhattacharya}}, \bibinfo {author} {\bibfnamefont {P.-L.}\ \bibnamefont {Giscard}},\ and\ \bibinfo {author} {\bibfnamefont {P.}~\bibnamefont {Meystre}},\ }\bibfield  {title} {\bibinfo {title} {Entangling the rovibrational modes of a macroscopic mirror using radiation pressure},\ }\href {https://doi.org/10.1103/PhysRevA.77.030303} {\bibfield  {journal} {\bibinfo  {journal} {Phys. Rev. A}\ }\textbf {\bibinfo {volume} {77}},\ \bibinfo {pages} {030303} (\bibinfo {year} {2008}{\natexlab{b}})}\BibitemShut {NoStop}%
	\bibitem [{\citenamefont {Rogers}\ \emph {et~al.}(2014)\citenamefont {Rogers}, \citenamefont {Gullo}, \citenamefont {Chiara}, \citenamefont {Palma},\ and\ \citenamefont {Paternostro}}]{Rogers_QMQM2014}%
	\BibitemOpen
	\bibfield  {author} {\bibinfo {author} {\bibfnamefont {B.}~\bibnamefont {Rogers}}, \bibinfo {author} {\bibfnamefont {N.~L.}\ \bibnamefont {Gullo}}, \bibinfo {author} {\bibfnamefont {G.~D.}\ \bibnamefont {Chiara}}, \bibinfo {author} {\bibfnamefont {G.~M.}\ \bibnamefont {Palma}},\ and\ \bibinfo {author} {\bibfnamefont {M.}~\bibnamefont {Paternostro}},\ }\bibfield  {title} {\bibinfo {title} {Hybrid optomechanics for quantum technologies},\ }\href {https://doi.org/doi:10.2478/qmetro-2014-0002} {\bibfield  {journal} {\bibinfo  {journal} {Quantum Measurements and Quantum Metrology}\ }\textbf {\bibinfo {volume} {2}},\ \bibinfo {pages} {000010247820140002} (\bibinfo {year} {2014})}\BibitemShut {NoStop}%
	\bibitem [{\citenamefont {Černotík}\ \emph {et~al.}(2019)\citenamefont {Černotík}, \citenamefont {Genes},\ and\ \citenamefont {Dantan}}]{Cernotik_QST2019}%
	\BibitemOpen
	\bibfield  {author} {\bibinfo {author} {\bibfnamefont {O.}~\bibnamefont {Černotík}}, \bibinfo {author} {\bibfnamefont {C.}~\bibnamefont {Genes}},\ and\ \bibinfo {author} {\bibfnamefont {A.}~\bibnamefont {Dantan}},\ }\bibfield  {title} {\bibinfo {title} {Interference effects in hybrid cavity optomechanics},\ }\href {https://doi.org/10.1088/2058-9565/aaf5a6} {\bibfield  {journal} {\bibinfo  {journal} {Quantum Science and Technology}\ }\textbf {\bibinfo {volume} {4}},\ \bibinfo {pages} {024002} (\bibinfo {year} {2019})}\BibitemShut {NoStop}%
	\bibitem [{\citenamefont {Barzanjeh}\ \emph {et~al.}(2022)\citenamefont {Barzanjeh}, \citenamefont {Xuereb}, \citenamefont {Gr{\"o}blacher}, \citenamefont {Paternostro}, \citenamefont {Regal},\ and\ \citenamefont {Weig}}]{Barzanjeh2022}%
	\BibitemOpen
	\bibfield  {author} {\bibinfo {author} {\bibfnamefont {S.}~\bibnamefont {Barzanjeh}}, \bibinfo {author} {\bibfnamefont {A.}~\bibnamefont {Xuereb}}, \bibinfo {author} {\bibfnamefont {S.}~\bibnamefont {Gr{\"o}blacher}}, \bibinfo {author} {\bibfnamefont {M.}~\bibnamefont {Paternostro}}, \bibinfo {author} {\bibfnamefont {C.~A.}\ \bibnamefont {Regal}},\ and\ \bibinfo {author} {\bibfnamefont {E.~M.}\ \bibnamefont {Weig}},\ }\bibfield  {title} {\bibinfo {title} {Optomechanics for quantum technologies},\ }\href {https://doi.org/10.1038/s41567-021-01402-0} {\bibfield  {journal} {\bibinfo  {journal} {Nature Physics}\ }\textbf {\bibinfo {volume} {18}},\ \bibinfo {pages} {15} (\bibinfo {year} {2022})}\BibitemShut {NoStop}%
	\bibitem [{\citenamefont {Regal}\ \emph {et~al.}(2008)\citenamefont {Regal}, \citenamefont {Teufel},\ and\ \citenamefont {Lehnert}}]{Regal_Nat2008}%
	\BibitemOpen
	\bibfield  {author} {\bibinfo {author} {\bibfnamefont {C.~A.}\ \bibnamefont {Regal}}, \bibinfo {author} {\bibfnamefont {J.~D.}\ \bibnamefont {Teufel}},\ and\ \bibinfo {author} {\bibfnamefont {K.~W.}\ \bibnamefont {Lehnert}},\ }\bibfield  {title} {\bibinfo {title} {Measuring nanomechanical motion with a microwave cavity interferometer},\ }\href {https://doi.org/10.1038/nphys974} {\bibfield  {journal} {\bibinfo  {journal} {Nature Physics}\ }\textbf {\bibinfo {volume} {4}},\ \bibinfo {pages} {555} (\bibinfo {year} {2008})}\BibitemShut {NoStop}%
	\bibitem [{\citenamefont {Chauhan}\ and\ \citenamefont {Biswas}(2016)}]{Chauhan_PRA2016}%
	\BibitemOpen
	\bibfield  {author} {\bibinfo {author} {\bibfnamefont {A.~K.}\ \bibnamefont {Chauhan}}\ and\ \bibinfo {author} {\bibfnamefont {A.}~\bibnamefont {Biswas}},\ }\bibfield  {title} {\bibinfo {title} {Atom-assisted quadrature squeezing of a mechanical oscillator inside a dispersive cavity},\ }\href {https://doi.org/10.1103/PhysRevA.94.023831} {\bibfield  {journal} {\bibinfo  {journal} {Phys. Rev. A}\ }\textbf {\bibinfo {volume} {94}},\ \bibinfo {pages} {023831} (\bibinfo {year} {2016})}\BibitemShut {NoStop}%
	\bibitem [{\citenamefont {Neukirch}\ \emph {et~al.}(2015)\citenamefont {Neukirch}, \citenamefont {von Haartman}, \citenamefont {Rosenholm},\ and\ \citenamefont {Nick~Vamivakas}}]{Neukirch_NatPhoton2015}%
	\BibitemOpen
	\bibfield  {author} {\bibinfo {author} {\bibfnamefont {L.~P.}\ \bibnamefont {Neukirch}}, \bibinfo {author} {\bibfnamefont {E.}~\bibnamefont {von Haartman}}, \bibinfo {author} {\bibfnamefont {J.~M.}\ \bibnamefont {Rosenholm}},\ and\ \bibinfo {author} {\bibfnamefont {A.}~\bibnamefont {Nick~Vamivakas}},\ }\bibfield  {title} {\bibinfo {title} {Multi-dimensional single-spin nano-optomechanics with a levitated nanodiamond},\ }\href {https://doi.org/10.1038/nphoton.2015.162} {\bibfield  {journal} {\bibinfo  {journal} {Nature Photonics}\ }\textbf {\bibinfo {volume} {9}},\ \bibinfo {pages} {653} (\bibinfo {year} {2015})}\BibitemShut {NoStop}%
	\bibitem [{\citenamefont {Morizot}\ \emph {et~al.}(2006)\citenamefont {Morizot}, \citenamefont {Colombe}, \citenamefont {Lorent}, \citenamefont {Perrin},\ and\ \citenamefont {Garraway}}]{Morizot_PRA2006}%
	\BibitemOpen
	\bibfield  {author} {\bibinfo {author} {\bibfnamefont {O.}~\bibnamefont {Morizot}}, \bibinfo {author} {\bibfnamefont {Y.}~\bibnamefont {Colombe}}, \bibinfo {author} {\bibfnamefont {V.}~\bibnamefont {Lorent}}, \bibinfo {author} {\bibfnamefont {H.}~\bibnamefont {Perrin}},\ and\ \bibinfo {author} {\bibfnamefont {B.~M.}\ \bibnamefont {Garraway}},\ }\bibfield  {title} {\bibinfo {title} {Ring trap for ultracold atoms},\ }\href {https://doi.org/10.1103/PhysRevA.74.023617} {\bibfield  {journal} {\bibinfo  {journal} {Phys. Rev. A}\ }\textbf {\bibinfo {volume} {74}},\ \bibinfo {pages} {023617} (\bibinfo {year} {2006})}\BibitemShut {NoStop}%
	\bibitem [{\citenamefont {Kanamoto}\ \emph {et~al.}(2003)\citenamefont {Kanamoto}, \citenamefont {Saito},\ and\ \citenamefont {Ueda}}]{PhysRevA.67.013608}%
	\BibitemOpen
	\bibfield  {author} {\bibinfo {author} {\bibfnamefont {R.}~\bibnamefont {Kanamoto}}, \bibinfo {author} {\bibfnamefont {H.}~\bibnamefont {Saito}},\ and\ \bibinfo {author} {\bibfnamefont {M.}~\bibnamefont {Ueda}},\ }\bibfield  {title} {\bibinfo {title} {Quantum phase transition in one-dimensional bose-einstein condensates with attractive interactions},\ }\href {https://doi.org/10.1103/PhysRevA.67.013608} {\bibfield  {journal} {\bibinfo  {journal} {Phys. Rev. A}\ }\textbf {\bibinfo {volume} {67}},\ \bibinfo {pages} {013608} (\bibinfo {year} {2003})}\BibitemShut {NoStop}%
	\bibitem [{\citenamefont {Schliesser}\ \emph {et~al.}(2006)\citenamefont {Schliesser}, \citenamefont {Del'Haye}, \citenamefont {Nooshi}, \citenamefont {Vahala},\ and\ \citenamefont {Kippenberg}}]{Schliesser_PRL2006}%
	\BibitemOpen
	\bibfield  {author} {\bibinfo {author} {\bibfnamefont {A.}~\bibnamefont {Schliesser}}, \bibinfo {author} {\bibfnamefont {P.}~\bibnamefont {Del'Haye}}, \bibinfo {author} {\bibfnamefont {N.}~\bibnamefont {Nooshi}}, \bibinfo {author} {\bibfnamefont {K.~J.}\ \bibnamefont {Vahala}},\ and\ \bibinfo {author} {\bibfnamefont {T.~J.}\ \bibnamefont {Kippenberg}},\ }\bibfield  {title} {\bibinfo {title} {Radiation pressure cooling of a micromechanical oscillator using dynamical backaction},\ }\href {https://doi.org/10.1103/PhysRevLett.97.243905} {\bibfield  {journal} {\bibinfo  {journal} {Phys. Rev. Lett.}\ }\textbf {\bibinfo {volume} {97}},\ \bibinfo {pages} {243905} (\bibinfo {year} {2006})}\BibitemShut {NoStop}%
	\bibitem [{\citenamefont {Arcizet}\ \emph {et~al.}(2006)\citenamefont {Arcizet}, \citenamefont {Cohadon}, \citenamefont {Briant}, \citenamefont {Pinard},\ and\ \citenamefont {Heidmann}}]{Arcizet_Nat2006}%
	\BibitemOpen
	\bibfield  {author} {\bibinfo {author} {\bibfnamefont {O.}~\bibnamefont {Arcizet}}, \bibinfo {author} {\bibfnamefont {P.-F.}\ \bibnamefont {Cohadon}}, \bibinfo {author} {\bibfnamefont {T.}~\bibnamefont {Briant}}, \bibinfo {author} {\bibfnamefont {M.}~\bibnamefont {Pinard}},\ and\ \bibinfo {author} {\bibfnamefont {A.}~\bibnamefont {Heidmann}},\ }\bibfield  {title} {\bibinfo {title} {Radiation-pressure cooling and optomechanical instability of a micromirror},\ }\href {https://doi.org/10.1038/nature05244} {\bibfield  {journal} {\bibinfo  {journal} {Nature}\ }\textbf {\bibinfo {volume} {444}},\ \bibinfo {pages} {71} (\bibinfo {year} {2006})}\BibitemShut {NoStop}%
	\bibitem [{\citenamefont {DeJesus}\ and\ \citenamefont {Kaufman}(1987)}]{Edmund_PRA1987}%
	\BibitemOpen
	\bibfield  {author} {\bibinfo {author} {\bibfnamefont {E.~X.}\ \bibnamefont {DeJesus}}\ and\ \bibinfo {author} {\bibfnamefont {C.}~\bibnamefont {Kaufman}},\ }\bibfield  {title} {\bibinfo {title} {Routh-hurwitz criterion in the examination of eigenvalues of a system of nonlinear ordinary differential equations},\ }\href {https://doi.org/10.1103/PhysRevA.35.5288} {\bibfield  {journal} {\bibinfo  {journal} {Phys. Rev. A}\ }\textbf {\bibinfo {volume} {35}},\ \bibinfo {pages} {5288} (\bibinfo {year} {1987})}\BibitemShut {NoStop}%
	\bibitem [{\citenamefont {Mancini}\ and\ \citenamefont {Tombesi}(1994{\natexlab{b}})}]{Mancini_PRA1994}%
	\BibitemOpen
	\bibfield  {author} {\bibinfo {author} {\bibfnamefont {S.}~\bibnamefont {Mancini}}\ and\ \bibinfo {author} {\bibfnamefont {P.}~\bibnamefont {Tombesi}},\ }\bibfield  {title} {\bibinfo {title} {Quantum noise reduction by radiation pressure},\ }\href {https://doi.org/10.1103/PhysRevA.49.4055} {\bibfield  {journal} {\bibinfo  {journal} {Phys. Rev. A}\ }\textbf {\bibinfo {volume} {49}},\ \bibinfo {pages} {4055} (\bibinfo {year} {1994}{\natexlab{b}})}\BibitemShut {NoStop}%
	\bibitem [{\citenamefont {Xu}\ and\ \citenamefont {Taylor}(2014)}]{Xunnong_PRA2014}%
	\BibitemOpen
	\bibfield  {author} {\bibinfo {author} {\bibfnamefont {X.}~\bibnamefont {Xu}}\ and\ \bibinfo {author} {\bibfnamefont {J.~M.}\ \bibnamefont {Taylor}},\ }\bibfield  {title} {\bibinfo {title} {Squeezing in a coupled two-mode optomechanical system for force sensing below the standard quantum limit},\ }\href {https://doi.org/10.1103/PhysRevA.90.043848} {\bibfield  {journal} {\bibinfo  {journal} {Phys. Rev. A}\ }\textbf {\bibinfo {volume} {90}},\ \bibinfo {pages} {043848} (\bibinfo {year} {2014})}\BibitemShut {NoStop}%
	\bibitem [{\citenamefont {Fabre}\ \emph {et~al.}(1994{\natexlab{b}})\citenamefont {Fabre}, \citenamefont {Pinard}, \citenamefont {Bourzeix}, \citenamefont {Heidmann}, \citenamefont {Giacobino},\ and\ \citenamefont {Reynaud}}]{Fabre_PRA1994}%
	\BibitemOpen
	\bibfield  {author} {\bibinfo {author} {\bibfnamefont {C.}~\bibnamefont {Fabre}}, \bibinfo {author} {\bibfnamefont {M.}~\bibnamefont {Pinard}}, \bibinfo {author} {\bibfnamefont {S.}~\bibnamefont {Bourzeix}}, \bibinfo {author} {\bibfnamefont {A.}~\bibnamefont {Heidmann}}, \bibinfo {author} {\bibfnamefont {E.}~\bibnamefont {Giacobino}},\ and\ \bibinfo {author} {\bibfnamefont {S.}~\bibnamefont {Reynaud}},\ }\bibfield  {title} {\bibinfo {title} {Quantum-noise reduction using a cavity with a movable mirror},\ }\href {https://doi.org/10.1103/PhysRevA.49.1337} {\bibfield  {journal} {\bibinfo  {journal} {Phys. Rev. A}\ }\textbf {\bibinfo {volume} {49}},\ \bibinfo {pages} {1337} (\bibinfo {year} {1994}{\natexlab{b}})}\BibitemShut {NoStop}%
	\bibitem [{\citenamefont {Militaru}\ \emph {et~al.}(2022)\citenamefont {Militaru}, \citenamefont {Rossi}, \citenamefont {Tebbenjohanns}, \citenamefont {Romero-Isart}, \citenamefont {Frimmer},\ and\ \citenamefont {Novotny}}]{Militaru_PRL2022}%
	\BibitemOpen
	\bibfield  {author} {\bibinfo {author} {\bibfnamefont {A.}~\bibnamefont {Militaru}}, \bibinfo {author} {\bibfnamefont {M.}~\bibnamefont {Rossi}}, \bibinfo {author} {\bibfnamefont {F.}~\bibnamefont {Tebbenjohanns}}, \bibinfo {author} {\bibfnamefont {O.}~\bibnamefont {Romero-Isart}}, \bibinfo {author} {\bibfnamefont {M.}~\bibnamefont {Frimmer}},\ and\ \bibinfo {author} {\bibfnamefont {L.}~\bibnamefont {Novotny}},\ }\bibfield  {title} {\bibinfo {title} {Ponderomotive squeezing of light by a levitated nanoparticle in free space},\ }\href {https://doi.org/10.1103/PhysRevLett.129.053602} {\bibfield  {journal} {\bibinfo  {journal} {Phys. Rev. Lett.}\ }\textbf {\bibinfo {volume} {129}},\ \bibinfo {pages} {053602} (\bibinfo {year} {2022})}\BibitemShut {NoStop}%
	\bibitem [{\citenamefont {Magrini}\ \emph {et~al.}(2022)\citenamefont {Magrini}, \citenamefont {Camarena-Ch\'avez}, \citenamefont {Bach}, \citenamefont {Johnson},\ and\ \citenamefont {Aspelmeyer}}]{Magrini_PRL2022}%
	\BibitemOpen
	\bibfield  {author} {\bibinfo {author} {\bibfnamefont {L.}~\bibnamefont {Magrini}}, \bibinfo {author} {\bibfnamefont {V.~A.}\ \bibnamefont {Camarena-Ch\'avez}}, \bibinfo {author} {\bibfnamefont {C.}~\bibnamefont {Bach}}, \bibinfo {author} {\bibfnamefont {A.}~\bibnamefont {Johnson}},\ and\ \bibinfo {author} {\bibfnamefont {M.}~\bibnamefont {Aspelmeyer}},\ }\bibfield  {title} {\bibinfo {title} {Squeezed light from a levitated nanoparticle at room temperature},\ }\href {https://doi.org/10.1103/PhysRevLett.129.053601} {\bibfield  {journal} {\bibinfo  {journal} {Phys. Rev. Lett.}\ }\textbf {\bibinfo {volume} {129}},\ \bibinfo {pages} {053601} (\bibinfo {year} {2022})}\BibitemShut {NoStop}%
	\bibitem [{\citenamefont {Ghobadi}\ \emph {et~al.}(2011)\citenamefont {Ghobadi}, \citenamefont {Bahrampour},\ and\ \citenamefont {Simon}}]{Ghobadi_PRA2011}%
	\BibitemOpen
	\bibfield  {author} {\bibinfo {author} {\bibfnamefont {R.}~\bibnamefont {Ghobadi}}, \bibinfo {author} {\bibfnamefont {A.~R.}\ \bibnamefont {Bahrampour}},\ and\ \bibinfo {author} {\bibfnamefont {C.}~\bibnamefont {Simon}},\ }\bibfield  {title} {\bibinfo {title} {Quantum optomechanics in the bistable regime},\ }\href {https://doi.org/10.1103/PhysRevA.84.033846} {\bibfield  {journal} {\bibinfo  {journal} {Phys. Rev. A}\ }\textbf {\bibinfo {volume} {84}},\ \bibinfo {pages} {033846} (\bibinfo {year} {2011})}\BibitemShut {NoStop}%
	\bibitem [{\citenamefont {Zhou}\ \emph {et~al.}(2024)\citenamefont {Zhou}, \citenamefont {Tang}, \citenamefont {Yin}, \citenamefont {Xia},\ and\ \citenamefont {Yin}}]{Zhou:24}%
	\BibitemOpen
	\bibfield  {author} {\bibinfo {author} {\bibfnamefont {J.}~\bibnamefont {Zhou}}, \bibinfo {author} {\bibfnamefont {J.}~\bibnamefont {Tang}}, \bibinfo {author} {\bibfnamefont {Y.}~\bibnamefont {Yin}}, \bibinfo {author} {\bibfnamefont {Y.}~\bibnamefont {Xia}},\ and\ \bibinfo {author} {\bibfnamefont {J.}~\bibnamefont {Yin}},\ }\bibfield  {title} {\bibinfo {title} {Fundamental probing limit on the high-order orbital angular momentum of light},\ }\href {https://doi.org/10.1364/OE.516620} {\bibfield  {journal} {\bibinfo  {journal} {Opt. Express}\ }\textbf {\bibinfo {volume} {32}},\ \bibinfo {pages} {5339} (\bibinfo {year} {2024})}\BibitemShut {NoStop}%
	\bibitem [{\citenamefont {Dezfouli}\ \emph {et~al.}(2022)\citenamefont {Dezfouli}, \citenamefont {Abramović}, \citenamefont {Rakić},\ and\ \citenamefont {Skenderović}}]{10.1063/5.0089735}%
	\BibitemOpen
	\bibfield  {author} {\bibinfo {author} {\bibfnamefont {A.~M.}\ \bibnamefont {Dezfouli}}, \bibinfo {author} {\bibfnamefont {D.}~\bibnamefont {Abramović}}, \bibinfo {author} {\bibfnamefont {M.}~\bibnamefont {Rakić}},\ and\ \bibinfo {author} {\bibfnamefont {H.}~\bibnamefont {Skenderović}},\ }\bibfield  {title} {\bibinfo {title} {{Detection of the orbital angular momentum state of light using sinusoidally shaped phase grating}},\ }\href {https://doi.org/10.1063/5.0089735} {\bibfield  {journal} {\bibinfo  {journal} {Applied Physics Letters}\ }\textbf {\bibinfo {volume} {120}},\ \bibinfo {pages} {191106} (\bibinfo {year} {2022})},\ \Eprint {https://arxiv.org/abs/https://pubs.aip.org/aip/apl/article-pdf/doi/10.1063/5.0089735/16479659/191106\_1\_online.pdf} {https://pubs.aip.org/aip/apl/article-pdf/doi/10.1063/5.0089735/16479659/191106\_1\_online.pdf} \BibitemShut {NoStop}%
	\bibitem [{\citenamefont {Ni}\ \emph {et~al.}(2022)\citenamefont {Ni}, \citenamefont {Liu}, \citenamefont {Yang}, \citenamefont {Tian}, \citenamefont {Hou}, \citenamefont {Shum},\ and\ \citenamefont {Chen}}]{Ni:22}%
	\BibitemOpen
	\bibfield  {author} {\bibinfo {author} {\bibfnamefont {W.}~\bibnamefont {Ni}}, \bibinfo {author} {\bibfnamefont {R.}~\bibnamefont {Liu}}, \bibinfo {author} {\bibfnamefont {C.}~\bibnamefont {Yang}}, \bibinfo {author} {\bibfnamefont {Y.}~\bibnamefont {Tian}}, \bibinfo {author} {\bibfnamefont {J.}~\bibnamefont {Hou}}, \bibinfo {author} {\bibfnamefont {P.~P.}\ \bibnamefont {Shum}},\ and\ \bibinfo {author} {\bibfnamefont {S.}~\bibnamefont {Chen}},\ }\bibfield  {title} {\bibinfo {title} {Annular phase grating-assisted recording of an ultrahigh-order optical orbital angular momentum},\ }\href {https://doi.org/10.1364/OE.473624} {\bibfield  {journal} {\bibinfo  {journal} {Opt. Express}\ }\textbf {\bibinfo {volume} {30}},\ \bibinfo {pages} {37526} (\bibinfo {year} {2022})}\BibitemShut {NoStop}%
	\bibitem [{\citenamefont {Pinnell}\ \emph {et~al.}(2020)\citenamefont {Pinnell}, \citenamefont {Rodríguez-Fajardo},\ and\ \citenamefont {Forbes}}]{Pinnell_2021}%
	\BibitemOpen
	\bibfield  {author} {\bibinfo {author} {\bibfnamefont {J.}~\bibnamefont {Pinnell}}, \bibinfo {author} {\bibfnamefont {V.}~\bibnamefont {Rodríguez-Fajardo}},\ and\ \bibinfo {author} {\bibfnamefont {A.}~\bibnamefont {Forbes}},\ }\bibfield  {title} {\bibinfo {title} {Probing the limits of orbital angular momentum generation and detection with spatial light modulators},\ }\href {https://doi.org/10.1088/2040-8986/abcd02} {\bibfield  {journal} {\bibinfo  {journal} {Journal of Optics}\ }\textbf {\bibinfo {volume} {23}},\ \bibinfo {pages} {015602} (\bibinfo {year} {2020})}\BibitemShut {NoStop}%
	\bibitem [{\citenamefont {He}\ \emph {et~al.}(2022)\citenamefont {He}, \citenamefont {Shen},\ and\ \citenamefont {Forbes}}]{He2022}%
	\BibitemOpen
	\bibfield  {author} {\bibinfo {author} {\bibfnamefont {C.}~\bibnamefont {He}}, \bibinfo {author} {\bibfnamefont {Y.}~\bibnamefont {Shen}},\ and\ \bibinfo {author} {\bibfnamefont {A.}~\bibnamefont {Forbes}},\ }\bibfield  {title} {\bibinfo {title} {Towards higher-dimensional structured light},\ }\href {https://doi.org/10.1038/s41377-022-00897-3} {\bibfield  {journal} {\bibinfo  {journal} {Light: Science {\&} Applications}\ }\textbf {\bibinfo {volume} {11}},\ \bibinfo {pages} {205} (\bibinfo {year} {2022})}\BibitemShut {NoStop}%
	\bibitem [{\citenamefont {Adesso}\ and\ \citenamefont {Illuminati}(2007)}]{Adesso_2007}%
	\BibitemOpen
	\bibfield  {author} {\bibinfo {author} {\bibfnamefont {G.}~\bibnamefont {Adesso}}\ and\ \bibinfo {author} {\bibfnamefont {F.}~\bibnamefont {Illuminati}},\ }\bibfield  {title} {\bibinfo {title} {Entanglement in continuous-variable systems: recent advances and current perspectives},\ }\href {https://doi.org/10.1088/1751-8113/40/28/S01} {\bibfield  {journal} {\bibinfo  {journal} {Journal of Physics A: Mathematical and Theoretical}\ }\textbf {\bibinfo {volume} {40}},\ \bibinfo {pages} {7821} (\bibinfo {year} {2007})}\BibitemShut {NoStop}%
	\bibitem [{\citenamefont {Adesso}\ and\ \citenamefont {Illuminati}(2006)}]{Adesso_2006}%
	\BibitemOpen
	\bibfield  {author} {\bibinfo {author} {\bibfnamefont {G.}~\bibnamefont {Adesso}}\ and\ \bibinfo {author} {\bibfnamefont {F.}~\bibnamefont {Illuminati}},\ }\bibfield  {title} {\bibinfo {title} {Continuous variable tangle, monogamy inequality, and entanglement sharing in gaussian states of continuous variable systems},\ }\href {https://doi.org/10.1088/1367-2630/8/1/015} {\bibfield  {journal} {\bibinfo  {journal} {New Journal of Physics}\ }\textbf {\bibinfo {volume} {8}},\ \bibinfo {pages} {15} (\bibinfo {year} {2006})}\BibitemShut {NoStop}%
	\bibitem [{\citenamefont {Li}\ \emph {et~al.}(2018)\citenamefont {Li}, \citenamefont {Zhu},\ and\ \citenamefont {Agarwal}}]{PhysRevLett.121.203601}%
	\BibitemOpen
	\bibfield  {author} {\bibinfo {author} {\bibfnamefont {J.}~\bibnamefont {Li}}, \bibinfo {author} {\bibfnamefont {S.-Y.}\ \bibnamefont {Zhu}},\ and\ \bibinfo {author} {\bibfnamefont {G.~S.}\ \bibnamefont {Agarwal}},\ }\bibfield  {title} {\bibinfo {title} {Magnon-photon-phonon entanglement in cavity magnomechanics},\ }\href {https://doi.org/10.1103/PhysRevLett.121.203601} {\bibfield  {journal} {\bibinfo  {journal} {Phys. Rev. Lett.}\ }\textbf {\bibinfo {volume} {121}},\ \bibinfo {pages} {203601} (\bibinfo {year} {2018})}\BibitemShut {NoStop}%
	\bibitem [{\citenamefont {Kippenberg}\ and\ \citenamefont {Vahala}(2007)}]{Kippenberg_OptExp2007}%
	\BibitemOpen
	\bibfield  {author} {\bibinfo {author} {\bibfnamefont {T.}~\bibnamefont {Kippenberg}}\ and\ \bibinfo {author} {\bibfnamefont {K.}~\bibnamefont {Vahala}},\ }\bibfield  {title} {\bibinfo {title} {Cavity opto-mechanics},\ }\href {https://doi.org/10.1364/OE.15.017172} {\bibfield  {journal} {\bibinfo  {journal} {Opt. Express}\ }\textbf {\bibinfo {volume} {15}},\ \bibinfo {pages} {17172} (\bibinfo {year} {2007})}\BibitemShut {NoStop}%
	\bibitem [{\citenamefont {Li}\ \emph {et~al.}(2021)\citenamefont {Li}, \citenamefont {Ou}, \citenamefont {Lei},\ and\ \citenamefont {Liu}}]{Li_NanoPhoto2021}%
	\BibitemOpen
	\bibfield  {author} {\bibinfo {author} {\bibfnamefont {B.-B.}\ \bibnamefont {Li}}, \bibinfo {author} {\bibfnamefont {L.}~\bibnamefont {Ou}}, \bibinfo {author} {\bibfnamefont {Y.}~\bibnamefont {Lei}},\ and\ \bibinfo {author} {\bibfnamefont {Y.-C.}\ \bibnamefont {Liu}},\ }\bibfield  {title} {\bibinfo {title} {Cavity optomechanical sensing},\ }\href {https://doi.org/doi:10.1515/nanoph-2021-0256} {\bibfield  {journal} {\bibinfo  {journal} {Nanophotonics}\ }\textbf {\bibinfo {volume} {10}},\ \bibinfo {pages} {2799} (\bibinfo {year} {2021})}\BibitemShut {NoStop}%
	\bibitem [{\citenamefont {Stannigel}\ \emph {et~al.}(2012)\citenamefont {Stannigel}, \citenamefont {Komar}, \citenamefont {Habraken}, \citenamefont {Bennett}, \citenamefont {Lukin}, \citenamefont {Zoller},\ and\ \citenamefont {Rabl}}]{Stannigel_PRL2012}%
	\BibitemOpen
	\bibfield  {author} {\bibinfo {author} {\bibfnamefont {K.}~\bibnamefont {Stannigel}}, \bibinfo {author} {\bibfnamefont {P.}~\bibnamefont {Komar}}, \bibinfo {author} {\bibfnamefont {S.~J.~M.}\ \bibnamefont {Habraken}}, \bibinfo {author} {\bibfnamefont {S.~D.}\ \bibnamefont {Bennett}}, \bibinfo {author} {\bibfnamefont {M.~D.}\ \bibnamefont {Lukin}}, \bibinfo {author} {\bibfnamefont {P.}~\bibnamefont {Zoller}},\ and\ \bibinfo {author} {\bibfnamefont {P.}~\bibnamefont {Rabl}},\ }\bibfield  {title} {\bibinfo {title} {Optomechanical quantum information processing with photons and phonons},\ }\href {https://doi.org/10.1103/PhysRevLett.109.013603} {\bibfield  {journal} {\bibinfo  {journal} {Phys. Rev. Lett.}\ }\textbf {\bibinfo {volume} {109}},\ \bibinfo {pages} {013603} (\bibinfo {year} {2012})}\BibitemShut {NoStop}%
\end{thebibliography}
%

\end{document}